\documentclass[nofootinbib,showpacs,preprint,aps]{revtex4}
\usepackage[reqno]{amsmath}
\usepackage{amsthm}
\usepackage{amssymb}
\usepackage{array}
\usepackage{longtable}
\usepackage{enumerate}
\usepackage{verbatim}
\usepackage{fancyhdr}
\usepackage{slashbox}
\usepackage{color}
\usepackage{graphicx}
\usepackage{bm}

\providecommand{\R}{\ensuremath{\mathbb{R}}}

\providecommand{\al}{\alpha}

\providecommand{\ga}{\gamma}
\providecommand{\Ga}{\Gamma}

\providecommand{\te}{\theta}
\providecommand{\ep}{\epsilon}
\providecommand{\sig}{\sigma}

\providecommand{\um}{\dfrac{1}{2}}
\DeclareMathOperator{\pRe}{Re} 
\DeclareMathOperator{\pIm}{Im} 
\DeclareMathOperator{\pArg}{Arg} 

\DeclareMathOperator{\erfc}{erfc}

\DeclareMathOperator{\Res}{Res}
\newcommand{\EL}[1]{\Biggl\{#1\Biggr\}} 
\newcommand{\GL}[1]{\biggl\{#1\biggr\}} 
\newcommand{\NL}[1]{\bigl\{#1\bigr\}}   

\newcommand{\EC}[1]{\Biggl[#1\Biggr]}   
\newcommand{\GC}[1]{\biggl[#1\biggr]}   
\newcommand{\MC}[1]{\Bigl[#1\Bigr]}     
\newcommand{\NC}[1]{\bigl[#1\bigr]}     

\newcommand{\GP}[1]{\biggl(#1\biggr)}   
\newcommand{\MP}[1]{\Bigl(#1\Bigr)}     
\newcommand{\NP}[1]{\bigl(#1\bigr)}     

\newcommand{\NB}[1]{\bigl|#1\bigr|}     
\newcommand{\GB}[1]{\biggl|#1\biggr|}   
\newcommand{\EB}[1]{\Biggl|#1\Biggr|}   
\newcommand{\En}[1]{\Biggr|_{#1}}   

\newcommand{\ODo}[1]{\frac{d}{d#1}}

\newcommand{\Do}[2]{\frac{d#1}{d#2}}

\newcommand{\intdef}[4]{\int_{#1}^{#2}{#3\,d#4}}

\newcommand{\intseminf}[2]{\int_{0}^{\,\infty}{#1\,d#2}}

\begin{document}
\setcounter{page}{1}
\title
{Different manifestations of S-matrix poles}
\author
{D. F. Ram\'irez Jim\'enez}
\email{df.ramirezj@uniandes.edu.co}
\author
{N. G. Kelkar}
\email{nkelkar@uniandes.edu.co}
\affiliation{ Departamento de Fisica, Universidad de los Andes,
Cra.1E No.18A-10, Santafe de Bogota, Colombia}
\begin{abstract}
Making use of the analytical properties of the $S$-matrix and a theorem 
of Mittag-Leffler, model independent non-relativistic expressions for cross sections 
in single channel elastic scattering, scattering phase shifts and 
survival probabilities of resonances are derived. 
Provided certain conditions are satisfied by the poles, the residues can also be 
estimated analytically. 
Considerations of the low energy behaviour of the $S$-matrix and cross sections 
reveal additional conditions on the residues of the poles appearing in the 
Mittag-Leffler expansions.  
The exact expressions for the resonant cross section and phase shift 
are shown to reduce to the commonly used Breit-Wigner formula plus corrections. 
The latter is shown to approach the exact result with the example of a meson 
and a baryon resonance. Finally, a comparison of the exact expressions with 
some realistic examples is presented. The calculations of survival probabilities 
in particular reveal the reason behind the non-observability of non-exponential 
decay.
\end{abstract}
\maketitle

\section{Introduction}
The occurrence of resonances and resonant phenomena is ubiquitous in nature. 
Indeed the only stable elementary objects in nature are the proton and electron. 
As a result, resonance production, propagation 
and its decay play an important role in almost all 
branches of physics. For example, the existence of a $^{12}$C resonance first 
predicted by Fred Hoyle \cite{hoyle54} and then found experimentally is crucial 
in order to explain the $^{12}$C abundance in the universe. Indeed, resonances 
appearing in nuclear processes play an important role in stellar nucleosynthesis
\cite{bertulanibook}. Many of the current research topics in particle and 
hadron physics also revolve around the searches \cite{etamesics} and understanding 
of different exotic unstable states  \cite{ruppandeef,Khemchandani}. 
Some decays such as the positronium for example, even play an 
important role in advances in medical technology \cite{jpet}. 
In the scattering of 
particles and nuclei, resonances usually show up as an enormous increase in the 
cross sections which are most commonly described by a Breit-Wigner form 
\cite{breitwigner} proposed several decades ago. 
Indeed, analytical forms of the Breit Wigner 
propagators were studied for decades along with their consistency with 
gauge invariance arguments (see \cite{Marek} and references therein). 
Though the latter works quite well 
in the case of most narrow resonances
\cite{blattweis,iliadis} (unstable states with large lifetimes), it 
should be remembered that it is only an approximate model. One can derive it for 
example from the argument that the exponential decay law (which is by itself an 
approximation) is encoded 
in the wave function, $|\Psi({\bf r}, t)|^2 = |\Psi({\bf r}, 0)|^2\, 
e^{-\Gamma t} \Rightarrow
\Psi({\bf r}, t) = \Psi({\bf r}, 0) e^{-\Gamma t/2}$. Expressing the wave function 
as a superposition of components having different energies, 
it is easy to obtain (after a Fourier transform)  
a Lorentzian or the Breit-Wigner form for the probability of finding the unstable 
state at an energy $E$. 

Characterization of particle resonances from data is in 
principle a very complex undertaking \cite{workmanplb} and there exist both model 
dependent approaches (with free parameters of the model 
fitted to data) or model independent 
approaches \cite{workmanprc} where the poles and residues 
in the expansion of the 
scattering amplitudes are fitted to data. Given the complexity of the analyses, 
model independent analytic expressions giving a physical insight into the 
time evolution and decay of resonances are not always available. In the present work, 
making use of the analytical properties of the 
$S$-matrix and a theorem of Mittag-Leffler (ML) \cite{MittagL}, an expansion of the 
$S$-matrix and thereby a general expression for the cross section depending on all 
the possible poles (bound, virtual, resonant) of the $S$-matrix is derived. 
The energy derivative of the phase shift (which, as shown by the Beth-Uhlenbeck 
formula \cite{BethUhl} corresponds to the density of states) is also 
obtained from the ML expansion and 
is used to derive analytical expressions for the 
survival amplitude within the Fock-Krylov method 
\cite{ournonexpo1,ournonexpo2}. 
The cross sections 
and phase shifts in single channel resonant elastic scattering derived using the 
above approach are shown to reduce 
to the commonly used Breit-Wigner forms plus corrections. 
We find that the derivation of all the above expressions near 
threshold requires special considerations and dedicate separate 
subsections to discuss the calculations of cross sections and 
survival probabilities for the resonances occurring close to 
threshold. The analytic $S$-matrix is a powerful tool which encodes the 
information of the bound, virtual, quasibound and resonant states through its 
poles. In what follows, we shall see its various manifestations in scattering 
and decay which enable us to 
obtain analytic expressions for the quantities mentioned above.

\section{Mittag-Leffler based analysis of cross sections}
Let us begin by recalling some basic facts \cite{Baz} 
on which the rest of this work will be based. Starting with 
the centre of mass momentum, $k$, in a two-body system where $k^2 = E$ (i.e., using 
$2m = \hbar = 1$): 
\begin{enumerate}[1)]
\item[i.]
If the Hamiltonian that describes an elastic scattering process is invariant under parity and time reversal, and if the partial wave $S$-matrix element, $S_l(k=k_o)=S_o$, then  
\begin{equation}\label{E2.6}
S_l(k=k_o^*)=\frac{1}{S_o^*},\quad S_l(k=-k_o^*)=S_o^*,\quad S_l(k=-k_o)=\frac{1}{S_o}.
\end{equation}
This property implies that if $k=k_o$ is a pole of $S_l(k)$, then 
$k=-k_o^*$ is also one, while $k=k_o^*$ and $k=-k_o$ are zeros of $S_l(k)$.
\item[ii.] Taking all the properties of the $S$ matrix into account \cite{Baz} 
allows us to characterize the poles and zeros of $S$ if the poles on the 
imaginary axis (both virtual and bound) and those in the third or fourth quadrant of the 
complex momentum $k$ plane are known.  
\end{enumerate}
The convention used for the poles and zeros of the $S$ matrix will be 
as follows: 
\begin{enumerate}[i)]
\item[iii.] $\NL{k_{ln}},\,n=1,2,\dotsc$ are simple poles of 
$S_l(k)$ in the fourth quadrant, i.e., $\pRe{k_{ln}}>0$ and 
$\pIm{k_{ln}}<0$ for all $n$.
\item[iv.] $\NL{i\zeta_{lm}}$ are simple poles of $S_l(k)$ on the imaginary axis. For $m=1,2,\dotsc$, $\zeta_{lm}>0$ and they represent bound states. For $m=-1,-2,\dotsc$, $\zeta_{lm}<0$ and they represent virtual states.
\end{enumerate}
The $S$ matrix will then have additional simple poles $\NL{-k_{ln}^*}$ and simple zeros at $\NL{-k_{ln}}$, $\NL{k_{ln}^*}$ and  $\NL{-i\zeta_{lm}}$. 
We remind that for mathematical simplicity, 
we set, $2m=\hbar = 1$ and hence $k^2 = E$. 
To clarify further the notation used, 
let us note that the resonance pole occurring at an energy $E_{lr}$ and width 
$\Gamma_{lr}$ in the $l^{th}$ partial wave is given 
for example by $k^2_{lr} = \epsilon_{lr} - i\Gamma_{lr}/2$ where 
$\epsilon_{lr} = E_{lr} - E_{th}$, with $E_{th}$ being the threshold energy
(or the sum of the masses of the decay products of the resonance).

\subsection{Expansion of the S-matrix}
A consequence of the above points (i) and (ii) is that 
$S_l(k)$ is a meromorphic function. Assuming that the 
poles of these functions are known, it is natural to ask if the information 
of the poles is sufficient to determine the function itself. 
The answer lies in a theorem of Mittag-Leffler (ML) \cite{MittagL}. 
\\
{\it If the only singularities of a meromorphic function $f(z)$ are the 
simple poles $z=a_1,a_2,\dotsc$ such that $|a_1| \le |a_2| \le ...$,  
with residues $b_1,b_2\dotsc$, respectively, and if 
$C_N$ is a circumference of radius $R_N$ which contains $N$ poles of the 
function $f(z)$ (and does not pass through any of the remaining poles), 
i.e., $|a_N|<R_N<|a_{N+1}|$, and in addition the function $f(z)$ satisfies $|f(z)|<M$, 
where $M$ is not dependent on $N$, then, when $N\to\infty$,
\begin{multline}\label{E2.7}
f(z)=f(0)+\lim_{N\to\infty}{\sum_{n=1}^{N}{b_{n}\GL{\frac{1}{z-a_n}+\frac{1}{a_n}}}}\\
+\lim_{N\to\infty}{\frac{z}{2\pi i}{\oint_{C_N}{\frac{f(\zeta)}{\zeta(\zeta-z)}\,d\zeta}}}
=f(0)+\sum_{n=1}^{\infty}{{\frac{b_nz}{a_n(z-a_n)}}}.
\end{multline}}
To derive the expressions for the cross sections, we begin by 
writing the Mittag-Leffler expansion of the $S$-matrix. In order to 
use the ML theorem, we must know the value of the function at the 
origin. To evaluate $S$ at $k=0$, we use the property \cite{Baz}:
$S_l(k)S_l(-k)=1$, 
as a consequence of which and using the definition of 
$S_l = \exp(2 i \delta_l)$, we get, 
$\delta_l(k)=-\delta_l(-k)$.
The phase shift, $\delta_l = 0$ for $k=0$ and hence 
$S_l(0)=1$.
With the convention mentioned in the beginning for the $S$-matrix poles and 
using the ML theorem, we obtain,   
\begin{equation}\label{E2.11}
S_l(k)=1+k\sum_n{\GC{\frac{b_{ln}}{k_{ln}(k-k_{ln})}-\frac{c_{ln}}{k_{ln}^*(k+k_{ln}^*)}}}+k\sum_m{\frac{d_{lm}}{i\zeta_{lm}(k-i\zeta_{lm})}},
\end{equation}
where,  
$b_{ln}=\Res{\NC{S_l(k),k=k_{ln}}}$,
$c_{ln}=\Res{\NC{S_l(k),k=-k_{ln}^*}}$ and 
$d_{lm}=\Res{\NC{S_l(k),k=i\zeta_{lm}}}$.

\subsubsection{Conditions imposed on the $S$-matrix expansion}
The analytic properties of the $S$-matrix impose the following conditions on 
\eqref{E2.11}:
\begin{enumerate}[i)]
\item {\it If $k$ is purely imaginary, $S_l(k)$ should be real:} 
If $k=i\ga$, with $\ga$ real, Eq. \eqref{E2.11} can be written as 
\begin{equation}\label{E2.15}
S_l(i\ga)=1+i\ga\sum_{n}{\GC{\frac{b_{ln}}{k_{ln}(i\ga-k_{ln})}-\frac{c_{ln}}{k_{ln}^*(i\ga+k_{ln}^*)}}}+i\ga\sum_m{\frac{d_{lm}}{i\zeta_{lm}(i\ga-i\zeta_{lm})}}.
\end{equation}
Taking the complex conjugate, 
\begin{align}
S_l^*(i\ga)
&=1-i\ga\sum_n{\GC{\frac{b_{ln}^*}{k_{ln}^*(-i\ga-k_{ln}^*)}-\frac{c_{ln}^*}{k_{ln}(-i\ga+k_{ln})}}}-i\ga\sum_m{\frac{d_{lm}^*}{-i\zeta_{lm}(-i\ga+i\zeta_{lm})}}\notag\\
&=1+i\ga\sum_n{\GC{\frac{b_{ln}^*}{k_{ln}^*(i\ga+k_{ln}^*)}-\frac{c_{ln}^*}{k_{ln}(i\ga-k_{ln})}}}+i\ga\sum_m{\frac{-d_{lm}^*}{i\zeta_{lm}(i\ga-i\zeta_{lm})}},\label{E2.16}
\end{align}
and comparing the two expressions, we deduce that 
$c_{ln}=-b_{ln}^*$ and $d_{lm}=-d_{lm}^*$. 
Considering the properties of the residues of the imaginary poles, we rewrite: 
\begin{equation}\label{E2.17}
D_{lm}\equiv\frac{d_{lm}}{i}=\frac{1}{i}\Res{\MC{S_l(k),k=i\zeta_{lm}}},
\end{equation}
and since $d_{lm}=-d_{lm}^*$, $D_{lm}$ is real. 
Finally, the expansion for the $S$-matrix can be written in the following 
form: 
\begin{equation}\label{E2.18}
\boxed{S_l(k)=1+k\sum_n{\GC{\frac{b_{ln}}{k_{ln}(k-k_{ln})}+\frac{b_{ln}^*}{k_{ln}^*(k+k_{ln}^*)}}}+k\sum_m{\frac{D_{lm}}{\zeta_{lm}(k-i\zeta_{lm})}}.}
\end{equation}
\item {\it $\NL{-k_{ln}}$, $\NL{k_{ln}^*}$ and  $\NL{-i\zeta_{lm}}$ 
are zeros of the $S$ matrix:} In this case, the expansion 
\eqref{E2.18} must satisfy the following equations:
\begin{multline}\label{E2.19}
S_l(-k_{ln})=1-k_{ln}\sum_p{\GC{\frac{b_{lp}}{k_{lp}(-k_{ln}-k_{lp})}+\frac{b_{lp}^*}{k_{lp}^*(-k_{ln}+k_{lp}^*)}}}\\-k_{ln}\sum_q{\frac{D_{lq}}{\zeta_{lq}(-k_{ln}-i\zeta_{lq})}}=0,
\end{multline}
\begin{multline}\label{E2.20}
S_l(k_{ln}^*)=1+k_{ln}^*\sum_p{\GC{\frac{b_{lp}}{k_{lp}(k_{ln}^*-k_{lp})}+\frac{b_{lp}^*}{k_{lp}^*(k_{ln}^*+k_{lp}^*)}}}\\+k_{ln}^*\sum_q{\frac{D_{lq}}{\zeta_{lq}(k_{ln}^*-i\zeta_{lq})}}=0,
\end{multline}
\begin{multline}\label{E2.21}
S_l(i\zeta_{lm})=1+i\zeta_{lm}\sum_{p}{\GC{\frac{b_{lp}}{k_{lp}(i\zeta_{lm}-k_{lp})}+\frac{b_{lp}^*}{k_{lp}^*(i\zeta_{lm}+k_{lp}^*)}}}\\+i\zeta_{lm}\sum_q{\frac{D_{lq}}{\zeta_{lq}(i\zeta_{lm}-i\zeta_{lq})}}=0.
\end{multline}
Eq. \eqref{E2.20} is the conjugate of \eqref{E2.19}. Therefore,
these conditions reduce to 
\begin{align}
1+k_{ln}\sum_p{\GC{\frac{b_{lp}}{k_{lp}(k_{ln}+k_{lp})}+\frac{b_{lp}^*}{k_{lp}^*(k_{ln}-k_{lp}^*)}}}+k_{ln}\sum_q{\frac{D_{lq}}{\zeta_{lq}(k_{ln}+i\zeta_{lq})}}&=0\label{E2.22},\\
1+i\zeta_{lm}\sum_p{\GC{\frac{b_{lp}}{k_{lp}(i\zeta_{lm}-k_{lp})}+\frac{b_{lp}^*}{k_{lp}^*(i\zeta_{lm}+k_{lp}^*)}}}+\zeta_{lm}\sum_q{\frac{D_{lq}}{\zeta_{lq}(\zeta_{lm}-\zeta_{lq})}}&=0.\label{E2.23}
\end{align}
\item {\it If $k$ is real, $S_l(k)$ must be unimodular} \cite{Baz}. 
This implies that the expansion \eqref{E2.18} must satisfy the condition:
$S_l(k)S_l^*(k)=1$.
The proof for this is given in Appendix A. 
\end{enumerate} 

\subsubsection{Residues in the $S$-matrix expansion}
To determine the $S$-matrix completely from the Mittag-Leffler expansion, 
we must now calculate the residues in (\ref{E2.18}). 
We mention here already that the complete expression for the residue of a single 
pole will be found to depend on all other (bound, virtual and resonant if any) poles
and difficult to use. Hence, we shall derive an approximate expression which can be 
used under certain conditions.
The $S$-matrix in 
literature and text books \cite{Newton} can be found to be written either 
in the context of potential scattering or otherwise. We write, 
\begin{equation}\label{E2.25}
S_l(k)=D(k)\prod_{n}{\frac{(k+k_{ln})(k-k_{ln}^*)}{(k-k_{ln})(k+k_{ln}^*)}}\prod_{m}{\frac{i\zeta_{lm}+k}{i\zeta_{lm}-k}},
\end{equation}
where (considering the properties of the $S$-matrix) the function $D(k)$ is 
unity for $k=0$ and describes the case of a potential of finite range $a$ 
or otherwise with the following form: 
\begin{equation}\label{E2.26}
D(k)=
\begin{cases}
1 & \text{General}\\
e^{-2iak} & \text{Potential picture}
\end{cases}
\end{equation}
The $S$-matrix from the above considerations satisfies the following: 
$S_l(0)=1$, $S$ is real when $k$ is purely imaginary and is modular 
when $k$ is real. The residue of $S_l(k)$ at $k=k_{ln}$ is calculated as, 
\begin{align}
b_n&=\Res{\NC{S_l(k),k=k_{ln}}}=\lim_{k\to k_{ln}}{(k-k_{ln})S_l(k)}\notag\\
&=2k_{ln}\frac{k_{ln}-k_{ln}^*}{k_{ln}+k_{ln}^*}\,D(k_{ln})
\prod_{p\neq n}{\frac{(k_{ln}+k_{lp})(k_{ln}-k_{lp}^*)}{(k_{ln}-k_{lp})(k_{ln}+k_{lp}^*)}}\prod_{m}{\frac{i\zeta_{lm}+k_{ln}}{i\zeta_{lm}-k_{ln}}}.\label{E2.27}
\end{align}
If we define the function, 
\begin{equation}\label{E2.28}
T_l^{(n)}(k)=D(k_{ln})\prod_{p\neq n}{\frac{(k+k_{lp})(k-k_{lp}^*)}{(k-k_{lp})(k+k_{lp}^*)}}\prod_{m}{\frac{i\zeta_{lm}+k}{i\zeta_{lm}-k}},
\end{equation}
the residue can be written as, 
\begin{equation}\label{E2.29}
b_{ln}=2ik_{ln}\frac{\pIm{k_{ln}}}{\pRe{k_{ln}}}
T_l^{(n)}(k_{ln})=2ik_{ln}\tan{\NP{\pArg{k_{ln}}}}T_l^{(n)}(k_{ln}).
\end{equation}
The function $T_l^{(n)}(k)$, by definition is unity at $k=0$ and hence we 
can write 
\begin{equation}\label{E2.30}
T_l^{(n)}(k)=1+G_l^{(n)}(k),
\end{equation}
where $G_l^{(n)}(k)$ is defined such that $G_l^{(n)}(k)=0$ for $k=0$.. 
Inserting \eqref{E2.30} in \eqref{E2.29}, 
\begin{equation}\label{E2.31}
b_{ln}=2ik_{ln}\tan{\NP{\pArg{k_{ln}}}}\NC{1+G_l^{(n)}(k_{ln})}.
\end{equation}
The above expression can 
be split into two parts: one which depends solely on the pole (of which we 
are evaluating the residue) and 
another which depends on the remaining poles. This is to say,
\begin{equation}\label{E2.32}
b_{ln}=b_{ln}^{(1)}+b_{ln}^{(2)},
\end{equation}
where 
\begin{align}
b_{ln}^{(1)}&=2ik_{ln}\tan{\NP{\pArg{k_{ln}}}},\label{E2.33}\\
b_{ln}^{(2)}&=2ik_{ln}\tan{\NP{\pArg{k_{ln}}}}G_l^{(n)}(k_{ln})\notag\\
&=2ik_{ln}\tan{\NP{\pArg{k_{ln}}}}\,D(k_{ln})\EL{\prod_{p\neq n}{\frac{(k_{ln}+k_{lp})(k_{ln}-k_{lp}^*)}{(k_{ln}-k_{lp})(k_{ln}+k_{lp}^*)}
\prod_{m}{\frac{i\zeta_{lm}+k_{ln}}{i\zeta_{lm}-k_{ln}}}-1}}.\label{E2.34}
\end{align}
Using the above decomposition, we can now try to get an approximate estimate of 
the residues depending only on the resonance poles. This is to say, we shall now 
verify if the second term is negligible and if not, under what conditions can it be 
neglected. Let us begin by writing, 
\begin{equation}\label{E2.35}
b_{ln}\approx b_{ln}^{(1)}=2ik_{ln}\tan{\NP{\pArg{k_{ln}}}}.
\end{equation}
Note that this estimate is independent of the interaction involved. 
In order to derive the condition mentioned above, we start by expanding 
the function $T_l^{(n)}(k)$ about $k=0$. In the case of potential scattering, 
from (\ref{E2.28}), we get,  
\begin{equation}\label{E2.36}
\Do{T_l^{(n)}(k)}{k}\En{k=0}=T_l^{(n)}(0)\ODo{k}\ln{\NC{T_l^{(n)}(k)}}\En{k=0}
=2i\GC{2\pIm{\sum_{p\neq n}{\frac{1}{k_{lp}}}}-\GP{a+\sum_m{\frac{1}{\zeta_{lm}}}}}.
\end{equation}
Thus 
\begin{equation}\label{E2.37}
T_l^{(n)}(k)=1+2ik\GC{2\pIm{\sum_{p\neq n}{\frac{1}{k_{lp}}}}-\GP{a+\sum_m{\frac{1}{\zeta_{lm}}}}}+O(k^2).
\end{equation}
The $S$-matrix residue can be written using \eqref{E2.29} as, 
\begin{equation}\label{E2.38}
b_{ln}=2ik_{ln}\tan{\NP{\pArg{k_{ln}}}}\EL{1+2ik_{ln}\GC{2\pIm{\sum_{p\neq n}{\frac{1}{k_{lp}}}}-\GP{a+\sum_m{\frac{1}{\zeta_{lm}}}}}}+O(k_{ln}^2).
\end{equation}
The above equation implies that the estimate of the residue (of the resonant pole) 
given by \eqref{E2.33}
is correct if, 
\begin{equation}\label{E2.39}
\EB{2ik_{ln}\GC{2\pIm{\sum_{p\neq n}{\frac{1}{k_{lp}}}}-\GP{a+\sum_m{\frac{1}{\zeta_{lm}}}}}}\ll1.
\end{equation}
This condition can be written in another way: 
\begin{equation}\label{E2.40}
\EB{2\pIm{\sum_{p\neq n}{\frac{1}{k_{lp}}}}-\GP{a+\sum_m{\frac{1}{\zeta_{lm}}}}}\ll\frac{1}{2|k_{ln}|}.
\end{equation}
If the above is satisfied, it is possible to use \eqref{E2.35}, 
otherwise we must use \eqref{E2.34}. In a later section, we shall use the 
expression \eqref{E2.35} in connection with the discussion of an isolated resonance. 
However, if one performs a complete analysis of the cross sections including all 
possible poles (not necessarily one resonant pole), then the above condition 
proves useful. In the absence of a potential picture, the condition
\eqref{E2.40} is obtained in a similar way.  
The condition in this case is, 
\begin{equation}\label{E2.41}
\EB{2\pIm{\sum_{p\neq n}{\frac{1}{k_{lp}}}}-{\sum_m{\frac{1}{\zeta_{lm}}}}}\ll\frac{1}{2|k_{ln}|}.
\end{equation}

We end this subsection by noting that the above expressions give us a means to determine 
the residues if the poles of the $S$-matrix are known. An extension of such an analysis 
to the multichannel case 
can be quite useful for partial wave analyses of cross sections. In \cite{workmanprc} 
for example, the authors perform expansions such as those in the present work. However, 
the residues are fitted as free parameters from experimental data. Analytical 
expressions such as the above can be used as consistency criteria on the fitted 
residues. 

\subsection{Single channel cross sections}
Having defined the $S$-matrix using the Mittag-Leffler expansion, we now 
apply this result to the calculation of the angle integrated cross section, 
$\sig$, which in terms of the scattering amplitude,  
$F(\te,k)$, can be written using the optical theorem as:
\begin{equation}\label{E2.42}
\sig=\frac{4\pi}{k}\pIm{f(0,k)}, 
\end{equation}
where 
\begin{equation}\label{E2.43}
f(\te,k)=\frac{1}{k}\sum_{l=0}^{\infty}{(2l+1)e^{\,i\delta_l}\sin{\delta_l}P_l(\cos{\te})}.
\end{equation} 
In terms of $S$, it can be written as,  
\begin{equation}\label{E2.44}
f(\te,k)=\sum_{l=0}^{\infty}{(2l+1)\frac{S_l(k)-1}{2ik}P_l(\cos{\te})}.
\end{equation}
Thus, 
\begin{equation}\label{E2.45}
\sig =-\frac{2\pi}{k^2}\sum_{l=0}^{\infty}{(2l+1)\pRe{\NC{S_l(k)-1}}}.
\end{equation}
Using \eqref{E2.18}, $\pRe{\NC{S_l(k)-1}}$ is given by, 
\begin{eqnarray}\label{E2.46}
2\pRe{\NC{S_l(k)-1}}
&=&\NC{S_l(k)-1}+\NC{S_l(k)-1}^*\\ \nonumber
&=&4\pRe{\sum_n{\frac{b_{ln}}{k_{ln}}\frac{k^2}{k^2-k_{ln}^2}}}+\sum_m{\frac{D_{lm}}{\zeta_{lm}}\frac{2k^2}{k^2+\zeta_{lm}^2}}.
\end{eqnarray}
Replacing \eqref{E2.46} in \eqref{E2.45} leads to, 
\begin{equation}\label{E2.47}
\boxed{
\sig =-{4\pi}\pRe{\sum_{n,l}{(2l+1)\frac{b_{ln}}{k_{ln}}\frac{1}{k^2-k_{ln}^2}}}-2\pi\sum_{m,l}{\frac{D_{lm}}{\zeta_{lm}}\frac{2l+1}{k^2+\zeta_{lm}^2}}} \,.
\end{equation}
The above equation gives the total cross section simply in terms of the poles of 
the $S$-matrix. Indeed, the cross section for a given value of $l$ is 
given by, 
\begin{equation}\label{E2.126}
\sigma^{(l)}=-{4\pi}(2l+1)\pRe{\sum_{n}{\frac{b_{ln}}{k_{ln}}
\frac{1}{k^2-k_{ln}^2}}}-2\pi\sum_{m}{\frac{D_{lm}}{\zeta_{lm}}
\frac{2l+1}{k^2+\zeta_{lm}^2}},
\end{equation}
where
\begin{equation}\label{E2.127}
\sig =\sum_{l=0}^{\infty}{\sigma^{(l)}}.
\end{equation}
\subsection{Threshold considerations: $S$-matrix and cross sections} 
The expressions for the $S$-matrix and total cross sections in the above 
discussions were derived without paying any special attention to the behaviour of 
these quantities near threshold. 
The behaviour of the $S$-matrix and total cross section for a given 
value of the orbital angular momentum $l$ can be found in standard text-books
to be given by \cite{Joachain}:
\begin{align}
S_l(k)&\sim1+2id_lk^{2l+1},\quad d_l\in\R\label{E2.48}\\
\sigma^{(l)}&\sim k^{4l}.\label{E2.49}
\end{align}
Though the expressions obtained for $S_l$ and $\sig^{(l)}$ 
as such are correct, if we perform a Taylor expansion of these expressions 
near threshold, we would not 
obtain the threshold behaviour mentioned above. 
The latter implies that if we wish to study the threshold behaviour explicitly using 
the Mittag-Leffler expansion, the poles and residues must satisfy certain conditions 
which have been derived explicitly in Appendix B. 

In case of the $S$-matrix, note that $S_l-1$ has a zero of order $2l+1$ at $k=0$. 
From \eqref{E2.121}, its poles and residues must satisfy: 
{\begin{equation}\label{E2.124}
\sum_n{\frac{b_{ln}}{k_{ln}^{\nu+1}}}+(-1)^\nu\sum_n{\frac{b_{ln}^*}{{(k_{ln}^*)}^{\nu+1}}}
+\sum_m{\frac{iD_{lm}}{(i\zeta_{lm})^{\nu+1}}}=0,\quad \nu=1,2,\dotsc,2l;
\end{equation}
and, from \eqref{E2.115}, $S_l(k)$ can be written as,
\begin{equation}\label{E2.125}
\frac{S_l(k)-1}{k^{2l+1}}=\sum_n{\frac{b_{ln}}{k_{ln}^{2l+1}(k-k_{ln})}}+\sum_n{\frac{b_{ln}^*}{(k_{ln}^*)^{2l+1}(k+k_{ln}^*)}}+\sum_m{\frac{iD_{lm}}{(i\zeta_{lm})^{2l+1}(k-i\zeta_{lm})}}.
\end{equation}
The partial cross section has poles at 
$k=\pm k_{ln},\pm k_{ln}^*,\pm i\zeta_{lm}$. The residues are given by, 
\begin{align}
\Res{\NP{\sig^{(l)},k=\pm k_{ln}}}&=\lim_{k\to\pm k_{ln}}\NP{k\mp k_{ln}}\sig^{(l)}=\mp\pi(2l+1)\frac{b_{ln}}{k_{ln}^2},\label{E2.128}\\
\Res{\NP{\sig^{(l)},k=\pm k_{ln}^*}}&=\lim_{k\to\pm k_{ln}^*}\NP{k\mp k_{ln}^*}\sig^{(l)}=\mp\pi(2l+1)\frac{b_{ln}^*}{{k_{ln}^*}^2},\label{E2.129}\\
\Res{\NP{\sig^{(l)},k=\pm i\zeta_{lm}}}&=\lim_{k\to\pm i\zeta_{lm}}\NP{k\mp i\zeta_{lm}}\sig^{(l)}=\mp\pi(2l+1)\frac{D_{lm}}{i\zeta_{lm}^2}.\label{E2.130}
\end{align}
The condition \eqref{E2.121} in this case is: 
\begin{equation}\label{E2.132}
\pRe{\GP{\sum_n{{\frac{b_{ln}}{k_{ln}^{2\mu+3}}}}}}
=\um\sum_m{(-1)^{s}\frac{D_{lm}}{\zeta_{lm}^{2\mu+3}}},\quad \mu=1,2,\dotsc,2l-1.
\end{equation}
and the cross section, using \eqref{E2.115}, is given as, 
\begin{equation}\label{E2.133}
\boxed{
\sig^{(l)}=-{4\pi}\,\frac{(2l+1)}{k^2}\, k^{2(2l+1)}\, 
\GP{
\pRe{\sum_{n}{\frac{b_{ln}}{k_{ln}^{4l+1}}\dfrac{1}{k^2-k_{ln}^2}}}+\um\sum_{m}{\frac{D_{lm}}{\zeta_{lm}^{4l+1}}\dfrac{1}{k^2+\zeta_{lm}^2}}}} \,.
\end{equation}
$S_l$ and $\sigma^{(l)}$, given by \eqref{E2.125} and \eqref{E2.133}, 
contain the correct threshold behaviour which is obtained only if the poles and 
residues satisfy the conditions \eqref{E2.124} and \eqref{E2.132}.

It can be seen from \eqref{E2.133} that the partial cross section diverges 
for large  $k$. This is expected since the above expressions 
were derived with threshold considerations. In other words, 
for a given value of $L$, all partial wave cross sections for $L +1$ and bigger can be 
neglected when $k \to 0$ (see also eqs. (\ref{E2.115})-(\ref{E2.117}) in the Appendix B).
Thus, 
\begin{equation}\label{E2.134}
\sig^{(l+1)}=O\MP{\sig^{(l)}},\quad k\to0.
\end{equation}
This further implies that the total cross section must be written in the following 
manner: 
\begin{equation}\label{E2.135}
\sig=\sum_{l=0}^{L}{\sig^{(l)}}+O\MP{\sig^{(L+1)}},\quad k\to0,
\end{equation}
where $L=0,1,2,\dotsc$ 
Comparing \eqref{E2.47} and \eqref{E2.135}, we can see that they lead to the 
same expression for $l=0$ and $k\to0$. Even if one considers more terms for $k\to0$,
both the expressions will display the same behaviour for  $k\to0$.  
Finally, we close by mentioning that for large values of $k$, independent of the 
value of $l$, one must use \eqref{E2.47}.

\section{Resonances} 
In the previous section, we obtained formulae for the cross sections 
in terms of all the poles and residues of the $S$-matrix which can be 
evaluated analytically. We shall now focus only on the resonant cross sections and 
obtain generalized expressions in terms of the pole values, i.e. the central value 
of the energy (or mass) and width of the resonance. It is gratifying to find that 
the generalized formula can be expressed in terms of the commonly used Breit-Wigner 
formula plus corrections. As in the previous section, we will derive the 
expressions with threshold considerations explicitly.

A word of caution is in order here. In what follows, we are going to 
approximate the general partial wave cross section formula (\ref{E2.126}) 
by disregarding all possible virtual and bound state poles as well as other 
resonances by picking only one term from the infinite sum. 
This is a reasonable approximation if all poles other than the resonant one 
under consideration lie much farther away and do not have large residues.
Unfortunately, this is not always the case. In realistic hadron spectroscopy, 
for example, the resonance widths are often of the same order of magnitude as the 
radial and/or angular splittings. Even if the latter is true, 
simplistic approximations such as the above provide an ease in mathematical 
manipulations and as we shall see below, also lead to some useful conclusions.

To start with, we note that by ``generalized expression" here, we mean the case of a 
resonance pole, $k_{lr}^2=\ep_{lr}-i\Ga_{lr}/2$, where 
$\Ga_{lr}/2\ep_{lr}\ll1$ is not necessarily true but $\Ga_{lr}/2\ep_{lr} < 1$. 
If $\Ga_{lr}/2\ep_{lr}\ll1$, however, we recover the Breit-Wigner resonance formula.
Recall that $\epsilon_{lr} = E_{lr} - E_{th}$, with $E_{th}$ being the threshold 
energy and $E_{lr}$ the real part of the pole in the complex energy plane. 
In what follows, we derive the resonant cross section formula depending on the 
parameter 
\begin{equation}\label{E2.66}
x_{lr}\equiv\frac{\Ga_{lr}}{2\ep_{lr}}\,, 
\end{equation}
which characterizes this ratio. Using the general expression (\ref{E2.47}) 
for the cross section derived earlier
and neglecting all other (such as the virtual and bound state) 
poles except the resonant one, 
we can write for an isolated resonance and a given orbital angular momentum $l$: 
\begin{equation}\label{E2.131a}
\sig^{(l)}=-4\pi(2l+1)\pRe{\GP{\frac{b_{lr}}{k_{lr}}\frac{1}{k^2-k_{lr}^2}}}=
-4\pi\frac{2l+1}{|k_{lr}|^2}\,\frac{\pRe{\NC{{b_{lr}k_{lr}^*}\NP{k^2-{k_{lr}^*}^2}}}}{\NB{k^2-{k_{lr}^*}^2}^2}.
\end{equation}
In terms of $k_{lr}^2=\ep_{lr}-i\Ga_{lr}/2$, we have, 
\begin{equation}\label{E2.132a}
\sig^{(l)}
=-4\pi\frac{2l+1}{|k_{lr}|^2}\,\frac{\pIm{\NP{b_{lr}k_{lr}^*}}}{\Ga_{lr}/2}\,
\cfrac{1+\cfrac{\pRe{\NP{b_{lr}k_{lr}^*}}}{\pIm{\NP{b_{lr}k_{lr}^*}}}\GP{\cfrac{k^2-\ep_{lr}}{\Ga_{lr}/2}}}{1+\GP{\cfrac{k^2-\ep_{lr}}{\Ga_{lr}/2}}^2}.
\end{equation}
From $b_{lr}=2ik_{lr}\tan{(\pArg{k_{lr}})}$, we find that 
$b_{lr}k_{lr}^*=2i|k_{lr}|^2\tan{(\pArg{k_{lr}})}$ with its real and imaginary 
parts given respectively by, 
$0$ and $2|k_{lr}|^2\tan{(\pArg{k_{lr}})}$. Therefore, 
\begin{equation}\label{E2.133a}
\frac{\pRe{\NP{b_{lr}k_{lr}^*}}}{\pIm{\NP{b_{lr}k_{lr}^*}}}=0,
\end{equation}
and
\begin{equation}\label{E2.132b}
\boxed{
\sig^{(l)}
=-4\pi\NP{2l+1}\,\frac{2\tan{(\pArg{k_{lr}})}}{\Ga_{lr}/2}\,
\cfrac{1}{1+\GP{\cfrac{k^2-\ep_{lr}}{\Ga_{lr}/2}}^2}}\,.
\end{equation}
On the other hand, the argument of $k_{lr}$ is:
\begin{equation}\label{E2.134a}
\pArg{\NP{k_{lr}}}=\pArg{\sqrt{\ep_{lr}-i\frac{\Ga_{lr}}{2}}}=-\um\arctan{\NP{x_{lr}}}.
\end{equation}
The identity $\tan{\frac{1}{2}z}=\csc{z}-\cot{z}$ allows us to write the tangent of
this argument as,
\begin{equation}\label{E2.135a}
\tan{(\pArg{k_{lr}})}={\frac{2\ep_{lr}}{\Ga_{lr}}}\MC{1-\sqrt{1+\NP{x_{lr}}^2}\,},
\end{equation}
and on expanding in powers of $\Ga_{lr}/2\ep_{lr}$,
\begin{equation}\label{E2.136a}
\tan{(\pArg{k_{lr}})}=-\um{x_{lr}}+\frac{1}{8}\NP{x_{lr}}^3+\cdots
\end{equation}
Substituting \eqref{E2.136a} in \eqref{E2.132b}, we find that 
\begin{equation}\label{E2.138a}
\sig^{(l)}=\frac{4\pi\NP{2l+1}}{\epsilon_{lr}}\frac{\NP{\Ga_{lr}/2}^2}{\NP{k^2-\ep_{lr}}^2+\NP{\Ga_{lr}/2}^2}\GC{1-\frac{1}{4}\GP{\frac{\Ga_{lr}}{2\ep_{lr}}}^2+
\frac{1}{8}\GP{\frac{\Ga_{lr}}{2\ep_{lr}}}^4-
\frac{5}{64}\GP{\frac{\Ga_{lr}}{2\ep_{lr}}}^6 + 
\dotsc}.
\end{equation}
where $\frac{\Ga_{lr}}{2\ep_{lr}}<1$ but $\Ga_{lr}/2\ep_{lr}\ll1$ is not necessarily 
true.
\begin{figure}[h]
\includegraphics[width=14cm,height=8cm]{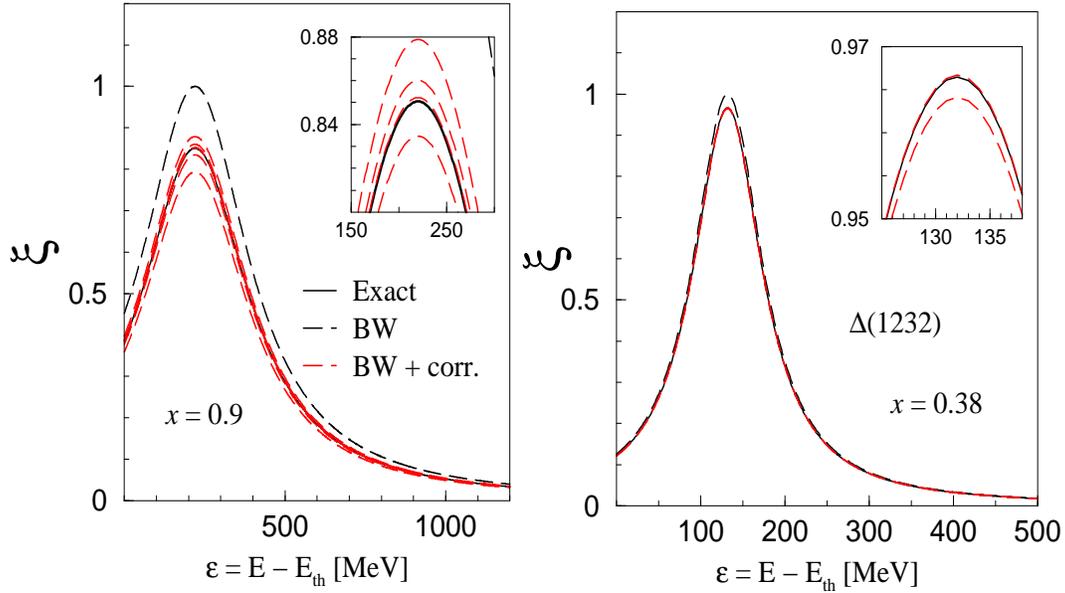}
\caption{Comparison of the exact cross section expression \eqref{E2.132b} 
(black solid line) with the 
commonly used Breit-Wigner cross section \eqref{E2.139a} (black dashed line). 
The cross section has been scaled by a factor to plot the dimensionless quantity, 
$\xi = \sigma^{(l)}/[4 \pi (2l +1) / \epsilon_{lr}]$. 
The red lines indicate the cross sections obtained after adding the corrections
given in \eqref{E2.138a} one by one to the Breit-Wigner cross section.  
The left panel shows the case of a broad resonance with the ratio 
$x = \Gamma_r/(2\epsilon_r)$ close to 1 and the right panel displays the case 
of the baryon resonance $\Delta$(1232) with a smaller $x$.
}
\end{figure}

In Fig. 1, we show a comparison of the exact expression \eqref{E2.132b} 
with the Breit-Wigner (BW) formula 
(the first term in \eqref{E2.138a}) for a broad ($\sigma$ meson) and a not so broad  
($\Delta$(1232)) resonance \cite{pdg}. 
We choose here an average mass and width of the $\sigma$ meson such that
$x= \Gamma_r/(2 \epsilon_r) < $ 1. The idea of this exercise is to simply demonstrate 
the comparison of the exact and the BW formula in the case of a broad and narrow 
resonance. In this respect, we must clarify to the reader that the intricate physics of 
the $\sigma$ meson \cite{tornqvist1,tornqvist2,isgur,tornqvist3,harada,tornqvist4,
eefbev,caprini,pelaez} 
cannot be contained in a naive single-pole description. This 
involves quark confinement, quark pair creation, coupled channels and an Adler zero
\cite{buggplb} which was shown to be crucial for the description of the $\sigma$ and 
$\kappa$ mesons. The choice of the $\sigma$ and $\Delta$(1232) may not be the best 
realistic choice \cite{nagels1,nagles2,nagles3} but serves the purpose of our 
exercise.

Coming back to Fig. 1, we see that the BW formula overestimates the cross sections. 
Addition of the correction terms in \eqref{E2.138a}, alternately decreases and 
increases the cross sections, bringing them closer to the exact result. 
The broad $\sigma$ meson which decays dominantly by $\sigma \to \pi \pi$ has a 
mass ($M_{\sigma}$) and width ($\Gamma_r$) of about 500 and 400 MeV 
respectively. Thus the value of $\epsilon_r = M_{\sigma} - 2 M_{\pi}$ is 220 MeV 
and $x= \Gamma_r/(2 \epsilon_r)$ is about 0.9. 
In case of the $\Delta$ resonance which decays by emitting a pion and a nucleon, 
the ratio, $x= \Gamma_r/(2 \epsilon_r)$ = 0.375.
A good agreement with the exact expression in case of the $\sigma$ is obtained after 
adding five corrections terms, whereas, for the $\Delta$, only two correction 
terms suffice. The insets in the figures display how the cross sections 
with BW+corrections approach the exact result.   
Since we are interested in comparing the exact results with the BW in the region 
around the peak of 
the resonance, we do not worry about the correct threshold behaviour of the cross 
section in Fig. 1. 

\subsection{Estimating the goodness of the Taylor expansion}
Let us begin with the following expression derived earlier: 
\begin{equation}\label{Eb1}
\sig^{(l)}=-4\pi(2l+1)\,\frac{2\tan{\NP{\pArg{k_{lr}}}}}{\Ga_{lr}/2}\,\cfrac{1}{1+\GP{\cfrac{k^2-\ep_{lr}}{\Ga_{lr}/2}}^2},
\end{equation}
where
\begin{equation}\label{Eb2}
\tan{\NP{\pArg{k_{lr}}}}=\frac{2\ep_{lr}}{\Ga_{lr}}\NC{1-\sqrt{1+x_{lr}^2}}=
\frac{1-\sqrt{1+x_{lr}^2}}{x_{lr}}.
\end{equation}
Substituting $z=x_{lr}^2$ and $\al=1/2$ in the expansion
\begin{equation}\label{Eb3}
\NP{1+z}^\al=\sum_{q=0}^{\infty}{\frac{\Ga\NP{\al+1}}{\Ga\NP{\al+1-q}}\,\frac{z^q}{q!}},
\end{equation}
we get:
\begin{equation}\label{Eb4}
\sqrt{1+x_{lr}^2}=\sum_{q=0}^{\infty}{\frac{\Ga\NP{\tfrac{3}{2}}}{\Ga\NP{\tfrac{3}{2}-q}}\,\frac{x_{lr}^{2q}}{q!}}=1+\um x_{lr}^2+\frac{\sqrt{\pi}}{2}\sum_{q=1}^{\infty}{\frac{1}{\Ga\NP{\tfrac{1}{2}-q}}\,\frac{x_{lr}^{2q+2}}{(q+1)!}}.
\end{equation}
From the properties of the Gamma function, the last equation can be written as
\begin{equation}\label{Eb5}
\sqrt{1+x_{lr}^2}=1+\frac{x_{lr}^2}{2}\sum_{q=0}^{\infty}{\frac{(-1)^q}{q+1}\,\frac{(2q)!}{(q!)^2}\,\GP{\frac{x_{lr}}{2}}^{2q}}.
\end{equation}
Hence, 
\begin{equation}\label{Eb6}
-\frac{2\tan{\NP{\pArg{k_{lr}}}}}{\Ga_{lr}/2}=\frac{1}{\ep_{lr}}\sum_{q=0}^{\infty}{\frac{(-1)^q}{q+1}\,\frac{(2q)!}{(q!)^2}\,\GP{\frac{x_{lr}}{2}}^{2q}},
\end{equation}
and the cross section is given by:
\begin{eqnarray}\label{Eb7}
\sig^{(l)}&=&\frac{4\pi(2l+1)}{\ep_{lr}}\,\frac{1}{1+K^2}\,\sum_{q=0}^{\infty}{\frac{(-1)^q}{q+1}\,\frac{(2q)!}{(q!)^2}\,\GP{\frac{x_{lr}}{2}}^{2q}} \nonumber \\
&=&\frac{4\pi(2l+1)}{\ep_{lr}}\,\frac{1}{1+K^2}\,\GP{1-\frac{1}{4}x_{lr}^2+\frac{1}{8}x_{lr}^4-\frac{5}{64}x_{lr}^6+\frac{7}{128}x_{lr}^8+\cdots},
\end{eqnarray}
where $K={\dfrac{k^2-\ep_{lr}}{\Ga_{lr}/2}}$. 
Let us now estimate the error in computing $\sig^{(l)}$ when we take $N-1$ terms of the 
last expansion. 
Since the series is alternating, the error is less than 
the absolute value of the next term omitted. 
Calling this error $\eta_N$ in units of $\frac{4\pi(2l+1)}{\ep_{lr}}$, 
\begin{equation}\label{Eb8}
\eta_N=\EB{\frac{\sig^{(l)}}{4\pi(2l+1)/\ep_{lr}}-\frac{1}{1+K^2}\,\sum_{q=0}^{N-1}{\frac{(-1)^q}{q+1}\,\frac{(2q)!}{(q!)^2}\,\GP{\frac{x_{lr}}{2}}^{2q}}}\leq\GC{\frac{1}{1+K^2}}\,\frac{1}{N+1}\,\frac{(2N)!}{(N!)^2}\,\GP{\frac{x_{lr}}{2}}^{2N}.
\end{equation}
The term in square brackets is always less that one for any value of $k$, hence, 
\begin{equation}\label{Eb9}
\eta_N\leq\frac{1}{N+1}\,\frac{(2N)!}{(N!)^2}\,\GP{\frac{x_{lr}}{2}}^{2N}.
\end{equation}
Applying this result to the examples discussed before, in case of the 
broad $\sigma$ meson we find that the error in calculating the cross section using 
5 correction terms is less than or equal to $1,4\times{10}^{-2}$ and 
the same in case of the $\Delta$ (1232) using only 2 terms is less than 
$2.6\times10^{-3}$.
To illustrate the usefulnes of the formula derived above, in the following table, we 
show the number of terms necessary in order to calculate the cross section using 
(\ref{Eb7}) such that the error is less than different powers of 10 shown in the 
table.

\begin{table}[ht]
\centering
\begin{tabular}{|c||*{6}{c|}}\cline{2-7}
\multicolumn{1}{c}{} & \multicolumn{6}{|c|}{$\eta_N\leq$}\\\hline
$x_{lr}$ & $10^{-1}$ & $10^{-2}$ & $10^{-3}$ & $10^{-4}$ & $10^{-5}$ & $10^{-6}$\\\hline
0.1 & 1 & 1 &  2 &  2 &  3&  3\\\hline
0.2 & 1 & 2 &  2 &  3 &  3&  4\\\hline
0.3 & 1 & 2 &  3 &  3 &  4&  5\\\hline
0.4 & 1 & 2 &  3 &  4 &  5&  6\\\hline
0.5 & 1 & 2 &  4 &  5 &  6&  8\\\hline
0.6 & 1 & 3 &  4 &  6 &  8& 10\\\hline
0.7 & 2 & 3 &  6 &  8 & 11& 14\\\hline
0.8 & 2 & 4 &  8 & 12 & 16& 20\\\hline
0.9 & 2 & 6 & 12 & 20 & 29& 38\\\hline
\end{tabular}
\caption{Number of terms necessary to calculate the cross section using 
the expansion \eqref{Eb7} with an error less than or equal to $\eta_N$.}
\end{table}

\subsection{Threshold considerations}
In order to incorporate the correct threshold behaviour in the cross section 
expression, we repeat the above exercise with the cross section 
expression \eqref{E2.133} 
valid close to threshold, i.e., starting with, 
\begin{equation}\label{E2.67}
\sig^{(l)}=-4\pi\pRe{\GL{(2l+1)\frac{b_{lr}}{k_{lr}}\GP{\frac{k}{k_{lr}}}^{4l}\dfrac{1}{k^2-k_{lr}^2}}}, 
\end{equation}
and replacing the residues as before and after some lengthy algebra, we find, 
\begin{equation}
\boxed{
\sig^{(l)}=-4\pi\,\frac{2l+1}{k^2}\, k^{2(2l+1)}\,
\frac{2\tan{\NP{\pArg{k_{lr}}}}}{\Gamma_{lr}/2}\,\frac{\pRe{\NP{{k_{lr}^*}^{4l}}}}{\NB{k_{lr}^2}^{4l}}\,\cfrac{1-\cfrac{\pIm{\NP{{k_{lr}^*}^{4l}}}}{\pRe{\NP{{k_{lr}^*}^{4l}}}}\GP{\cfrac{k^2-\ep_{lr}}{\Ga_{lr}/2}}}{1+\GP{\cfrac{k^2-\ep_{lr}}{\Ga_{lr}/2}}^2}}\, .
\label{E2.68}
\end{equation}
Once again, performing expansions in terms of the variable $x_{lr}$, one can show, 
\begin{multline}\label{E2.77}
\sig^{(l)}=4 \pi \frac{(2l+1)}{k^2}\,\frac{k^{2(2l+1)}}{\ep_{lr}^{2l}}\,
\Biggl[\frac{{\Ga_{lr}/2}}{\NP{k^2-\ep_{lr}}^2+\NP{\Ga_{lr}/2}^2}\,\GP{\frac{\Ga_{lr}}{2\ep_{lr}}}\\
-\frac{2l\NP{k^2-\ep_{lr}}}{\NP{k^2-\ep_{lr}}^2+\NP{\Ga_{lr}/2}^2}\,\GP{\frac{\Ga_{lr}}{2\ep_{lr}}}^2-\frac{8l^2+4l+1}{4}\,\frac{{\Ga_{lr}/2}}{\NP{k^2-\ep_{lr}}^2+\NP{\Ga_{lr}/2}^2}\,\GP{\frac{\Ga_{lr}}{2\ep_{lr}}}^3+\cdots\Biggr]
\end{multline}

Finally, some comments regarding \eqref{E2.138a} and \eqref{E2.77} are in order.
\begin{enumerate}[1)]
\item Considering the first term in the expansion of $\sig^{(l)}$ 
in Eq. \eqref{E2.138a}, 
i.e. for the case when $\Ga_{lr}/2\ep_{lr}\ll1$:
\begin{equation}\label{E2.139a}
\sig^{(l)}=\frac{4\pi\NP{2l+1}}{\epsilon_{lr}}\frac{\NP{\Ga_{lr}/2}^2}{\NP{k^2-\ep_{lr}}^2+\NP{\Ga_{lr}/2}^2} 
\end{equation}
and we obtain the standard Breit-Wigner distribution for the total cross section. 
\item The first term of $\sig^{(l)}$ given by \eqref{E2.77}:
\[
\sig^{(l)}=
\frac{4\pi(2l+1)}{k^2}\GP{\frac{k^2}{\pRe{k_{lr}^2}}}^{2l+1}\frac{\NP{\Ga_{lr}/2}^2}{\NP{k^2-\ep_{lr}}^2+\NP{\Ga_{lr}/2}^2} \,,
\]
explicitly displays the threshold behaviour which is similar for example 
to that used in the partial wave analysis of baryon resonances \cite{arndt3}. 
The above equation reduces to Eq. \eqref{E2.138a} for the $l = 0$ case. 
It is gratifying to recover the commonly used Breit-Wigner distributions for 
narrow resonances from the generalized expressions \eqref{E2.132b} and \eqref{E2.68}. 

\end{enumerate}

\section{Energy derivative of the phase shift} 
In the Fock-Krylov method \cite{Fock1}, 
the survival amplitude, $A_l(t)$, of an unstable state is given by the 
Fourier transform of the density of states $\rho_l(E)$ (usually taken to be of a 
Lorentzian form) with a threshold factor and a form factor $f(E)$ to ensure 
that the energy distribution $\rho_l(E) \to 0$ for large $E$. The latter, with 
the use of the Beth-Uhlenbeck formula for the difference in the density of
states with and without interaction, can be written in terms of the energy 
derivative of the phase shift ($\delta_l$) \cite{BethUhl,ournonexpo1}. 
While calculating the second virial coefficients $B$, $C$ 
in the equation of an ideal gas, 
$P V = R T (1 + \, B/V \,+\, C/V^2 \, + ...)$, Beth and Uhlenbeck found that 
the difference in the density of states (of scattered particles) with 
interaction $dn_l(E)/dE$ and $dn_l^{(0)}(E)/dE$ without interaction is 
\begin{equation}\label{bethuhlenformula} 
\rho_l(E) = \frac{dn_l(E)}{dE} \, - \, \frac{dn_l^{(0)}(E)}{dE} = 
\frac{2l + 1}{\pi} \, \frac{d\delta_l(E)}{dE} \,.
\end{equation}   
In resonant scattering, this is the density of states of a resonance 
(in terms of the decay products) \cite{KelkarAIP} and is useful in 
writing the survival amplitude of the resonance as, 
\begin{equation}\label{E2.1}
A_l(t)=\intseminf{f(E)\GP{\Do{\delta_l}{E}}e^{-iEt}}{E}.
\end{equation}
Here, $E = E_{CM} - E_{th}$, where $E_{CM}$ is the energy available in 
the centre of mass of the decay products and hence the lower limit of 
integration corresponding to $E_{CM} = E_{th}$ is $E=0$. 
The form factor $f(E)$ is commonly taken to be, $f(E)=e^{-cE},\quad c>0$
\cite{fonda}. 
The amplitude can further be expressed as, 
\begin{equation}\label{E2.3}
A_l(t)=\intseminf{\GP{\Do{\delta_l}{E}}e^{-(c+it)E}}{E}=\intseminf{\GP{\Do{\delta_l}{E}}e^{-sE}}{E},
\end{equation}
where $s\equiv{c+it}$, and $\pRe{s}>0$.} The probability 
amplitude written in this manner is the Laplace transform of 
${d\delta_l}/{dE}$. The energy derivative of the phase shift in elastic 
scattering was also shown to represent the time delay 
introduced in the scattering process due to the propagation of a resonance, by 
Wigner \cite{wigner} in 1955. 

We shall now calculate the derivative of the 
phase shift with respect to $k$ (here $k^2={E}$) and 
eventually relate it to the energy derivative. In terms of the $S$-matrix,
$S_l(k)=e^{\,2i\delta_l(k)}$ and   
the phase shift derivative is given by,  
\begin{equation}\label{E2.5}
\Do{\delta_l}{k}\equiv h_l(k)=\frac{1}{2i}\frac{1}{S_l(k)}\Do{S_l(k)}{k}.
\end{equation} 
In order to evaluate \eqref{E2.3}, knowledge of the analytical properties 
of ${d\delta_l}/{dk}$, 
which depend on the properties of $S_l(k)$ are required. The poles of $h_l(k)$ 
correspond either to the zeros or poles of $S_l(k)$. With $S_l(k)$ being a 
meromorphic function (as observed with the points in the beginning), we shall now 
attempt a Mittag-Leffler expansion of $h_l(k)$. To this end, let us first evaluate 
the residues of $h_l(k)$.  

\subsection{Residues of the phase shift derivative}
The expansion of the derivative of the phase shift, 
$h_l(k)$ in \eqref{E2.5}, can be done with the knowledge of the 
residues of the corresponding poles. 
If $k=P$ is a pole of $S_l(k)$, using the Laurent theorem, 
the $S$-matrix can be written as, 
\begin{equation}\label{E2.78}
S_l(k)=\frac{B}{k-P}+F_l(k),
\end{equation}
where $B$ is the residue of $S_l(k)$ at the given pole and 
$F_l(k)$ is an analytic function in the vicinity of $k=P$. 
$h_l(k)$ can be written as follows:
\begin{equation}\label{E2.79}
h_l(k)=\frac{1}{2i}\frac{1}{S_l(k)}\Do{S_l(k)}{k}
=\frac{1}{2i}\,\cfrac{-\cfrac{B}{(k-P)^2}+F_l'(k)}{\cfrac{B}{k-P}+F_l(k)}
=\frac{i}{2}\,\frac{1}{k-P}
\,\cfrac{1-\cfrac{(k-P)^2}{B}F_l'(k)}
{1+\cfrac{k-P}{B}F_l(k)}\,\,.
\end{equation}
The rightmost ratio in the above expression is analytic at $k=P$ and is 
unity for $k=P$. On performing a Taylor series expansion about $k=P$, 
the first term in the expansion does not change the coefficient $i/2$ 
(the residue of $h_l(k)$ at $k=P$) of $(k-P)^{-1}$. 

Let us now assume that $k=Z$ is a simple zero of $S_l(k)$. 
In this case, we can write $S_l(k)$ as, 
\begin{equation}\label{E2.80}
S_l(k)=(k-Z)\NC{b_1+b_2(k-Z)+\cdots}
\end{equation}
such that, $h_l(k)$ is given by, 
\begin{equation}\label{E2.81}
h_l(k)=\frac{1}{2i}\frac{1}{S_l(k)}\Do{S_l(k)}{k}
=\frac{1}{2i}\,\frac{1}{k-Z}\,\frac{b_1+2b_2(k-Z)+\cdots}{{b_1+b_2(k-Z)+\cdots}}
\end{equation}
Once again, the rightmost ratio is analytic at $k=Z$ and its value at this 
point is 1. Expanding once again about $k=Z$, the coefficient of 
$(k-Z)^{-1}$, namely, $-i/2$, which is the residue of $h_l(k)$ at $k=Z$ is 
not altered. To summarize the above:

{\it If $k=P$ is a pole of the $S$-matrix, the residue of $h_l(k)={d\delta_l}/{dk}$ at $k=P$ is:
\begin{equation}\label{E2.82}
\Res{\GP{\Do{\delta_l}{k},P}}=\lim_{k\to k_{ln}}{(k-P)\Do{\delta_l}{k}}=+\frac{i}{2},
\end{equation} 
and if $k=Z$ is a zero of the $S$-matrix, the residue of $h_l(k)={d\delta_l}/{dk}$ at $k=Z$ is:
\begin{equation}\label{E2.82a}
\Res{\GP{\Do{\delta_l}{k},Z}}=\lim_{k\to k_{ln}}{(k-Z)\Do{\delta_l}{k}}=-\frac{i}{2},
\end{equation}}

Note that these residues are independent of the location and the kind of 
poles as well as the potential involved.
\subsection{Mittag-Leffler expansion of the phase shift derivative}
We are now in a position to write the Mittag-Leffler expansion of the phase shift 
derivative. In what follows, we shall neglect virtual and bound state poles.
Considering the properties of $h_l(k)$ discussed so far, if 
${k_{ln}}$ are poles of the $S$ matrix in the fourth quadrant, then they are also
poles of $h_l(k)$, just as ${-k_{ln}}$ and ${\pm k_{ln}^*}$. 
Before applying the Mittag-Leffler theorem, we recall that in the 
vicinity of $k=0$, $h_l(k)={d\delta_l}/{dk}\sim A_lk^{2l}$, 
where $A_l$, for example, in case of potential scattering, 
is a constant given by \cite{Joachain}  
\begin{equation}\label{E2.83}
A_l=a^{2l+1}\,\frac{a^{2l+1}}{(2l+1)!!(2l-1)!!}\,\frac{l-a\hat{\ga}_l}{l+1+a\hat{\ga}_l},
\end{equation}
with $a$ being the range of the potential and $\hat{\ga}_l$ 
the derivative of the radial wave function at $r=a$. 
As a result, at $k=0$ the latter is $A_l\delta_{l0}$, 
and the Mittag-Leffler theorem leads us to 
\begin{eqnarray}\label{E2.84}
h_l(k)&=&A_l\delta_{l0}+\sum_n{\frac{i}{2}\,\frac{k}{k_{ln}}\,\GP{\frac{1}{k-k_{ln}}+\frac{1}{k+k_{ln}}}}-\sum_n{\frac{i}{2}\,\frac{k}{k_{ln}^*}\,\GP{\frac{1}{k-k_{ln}^*}+\frac{1}{k+k_{ln}^*}}} \\ 
\nonumber
&=&A_l\delta_{l0}-2k^2\pIm{\sum_n{\frac{1}{k_{ln}(k^2-k_{ln}^2)}}}\, .
\end{eqnarray}
The derivative with respect to energy is hence given by
\begin{equation}\label{E2.85}
\boxed{
\Do{\delta_l}{E}
=\Do{\delta_l}{k}\Do{k}{E}=\frac{h_l(\sqrt{E})}{2\sqrt{E}}
=\frac{A_l}{2}\delta_{l0}\,E^{-1/2}-\pIm{\sum_n{\frac{\sqrt{E}}{k_{ln}(E-k_{ln}^2)}}}}\, .
\end{equation}

A brief discussion of the physical meaning of the above relation at this 
point is in order. Let us recall that the energy derivative of the phase shift 
in single channel elastic scattering also represents Wigner's time delay 
\cite{wigner} due to interaction. This ``phase" time delay, in the $l=0$ 
case displays a singularity near threshold. If one considers the phase 
time in one dimensional tunneling, the singularity can be shown to 
arise due to the interference between the incident and reflected waves 
in front of the barrier. Subtraction of the singular term 
leads to the dwell time delay which was shown in \cite{Kelkar3} to reproduce
the correct behaviour near threshold. Apart from this, in \cite{Kelkar4} 
the authors demonstrated that the dwell time indeed has a physical meaning 
and gives the half lives of radioactive nuclei. 
In the above expression for the energy derivative of the phase shift, we can see 
using Eq. \eqref{E2.48}, that the interference term is proportional to 
\[
\frac{\pIm{S_0(k)}}{k}\Do{k}{E}=\frac{d_0}{2\sqrt{E}}=\um d_0E^{-1/2}.
\]
Since the energy derivative of the phase shift also corresponds to the density of 
states, \eqref{E2.85} is expected to give the correct density for $S$-waves 
only after a subtraction of the singular term. Thus, the corrected density is given 
by 
\begin{equation}\label{E2.86}
\boxed{
\biggl ( \Do{\delta_l}{E} \biggr )_C=-\pIm{\sum_n{\frac{\sqrt{E}}
{k_{ln}(E-k_{ln}^2)}}}.}
\end{equation}

Integrating the above expression, Eq. \eqref{E2.86} for the derivative of the 
phase shift, we obtain the phase shift: 
\begin{equation}\label{E2.89}
\delta_l(E)=\intdef{0}{E}
 {\biggl ( \Do{\delta_l}{E} \biggr )_C}{E} 
=\pIm{\sum_n{\frac{1}{k_{ln}}\intdef{0}{E}{\frac{E^{1/2}}{k_{ln}^2-E}}{E}}}.
\end{equation}
With a change of variable $E=k_{ln}^2\tanh^2{z}$ and integrating, 
\begin{equation}\label{E2.90}
\boxed{
\delta_l(E)=2\pIm\sum_n{\EC{\tanh^{-1}{\GP{\frac{\sqrt{E}}{k_{ln}}}}-{\frac{\sqrt{E}}{k_{ln}}}}}}.
\end{equation}

To end this section, we investigate the behaviour of the phase shift derivative in case 
of an isolated resonance. Given a pole at 
$E_{lr}=k_{lr}^2=\ep_{lr}-i\Ga_{lr}/2$, we begin from \eqref{E2.89} by writing it as:
\begin{equation}\label{E2.91}
\Do{\delta_l}{E}
=\pIm{{\frac{1}{k_{lr}}\GP{\frac{E^{1/2}}{k_{lr}^2-E}}}}.
\end{equation}
Expanding $E^{1/2}$ about $E=k_{lr}^2$ leads us to
\begin{equation}\label{E2.92}
\Do{\delta_l}{E}=\pIm{{{\frac{1}{k_{lr}^2-E}}}}+\cdots
\end{equation}
Evaluating the imaginary part, 
\begin{equation}\label{E2.93}
\Do{\delta_l}{E}=\frac{\pIm{\NC{(k_{lr}^*)^2-E}}}{|k_{lr}^2-E|^2}+\cdots=\frac{\Ga_{lr}/2}{(E-\ep_{lr})^2+(\Ga_{lr}/2)^2}+\cdots
\end{equation}
and we notice that the phase shift derivative around an isolated resonance pole 
can be described by a Breit-Wigner distribution plus corrections. 
Integrating Eq. 
\eqref{E2.93} with respect to $E$, we get 
\begin{equation}\label{E2.94}
\delta_l(E)=
\arctan{\GP{\cfrac{\Ga_{lr}/2}{\ep_{lr}-E}}}+\cdots
\end{equation}
which is once again the standard expression for a Breit-Wigner phase shift 
as found in text books \cite{Joachain}. 
Eqs \eqref{E2.93} and \eqref{E2.94} are valid in the vicinity of an isolated resonance. 
\subsection{Threshold behaviour of the phase shift derivative}
Extending the argument of the interference term discussed in the previous subsection 
for non-zero orbital angular momenta, such a term is proportional to 
\cite{Kelkar3,winful}, 
\[
\frac{\pIm{S_l(k)}}{k}\Do{k}{E}=\um d_lE^{l-1/2}.
\]
This means if we perform a Taylor series expansion of $\Do{\delta_l}{E}$ about $E=0$:
\[
\Do{\delta_l}{E}=d_lE^{l-1/2}\NP{1+a_1E+a_2E^2+\cdots},
\] 
the first term should be eliminated in order to obtain the correct behaviour of the 
phase shift derivative at low energies. Indeed, as expected, we do obtain 
the correct threshold behaviour  $E^{l+1/2}$ \cite{ournonexpo1,fonda}.
As derived in a previous section, the poles and residues of ${d\delta_l}/{dk}$ 
must satisfy the condition \eqref{E2.121}:
\begin{eqnarray}\label{E2.87}
&&\sum_n{\frac{i/2}{k_{ln}^{\eta+1}}}+\sum_n{\frac{i/2}{(-k_{ln}^*)^{\eta+1}}}-
\sum_n{\frac{i/2}{(-k_{ln})^{\eta+1}}}-\sum_n{\frac{i/2}{(k_{ln}^*)^{\eta+1}}}\\ \nonumber
&=&2i\sum_n{\NC{1+(-1)^\eta}\pIm{\GP{\frac{1}{k_{ln}^{\eta+1}}}}}\\ \nonumber
&=&0, \quad \eta=1,2,\dotsc,2l+1.
\end{eqnarray}
For odd $\eta$, we obtain a null condition from \eqref{E2.87}. 
Substituting $\eta\to2\eta$ in the previous equation, 
\begin{equation}\label{E2.88}
\pIm{\sum_n{\GP{\frac{1}{k_{ln}^{2\eta+1}}}}}=0,\quad \eta=1,2,\dotsc,l.
\end{equation}
Thus, the expression for the phase shift derivative near threshold is:
\begin{eqnarray}\label{E2.89a}
\Do{\delta_l}{E}
&=&\Do{\delta_l}{k}\Do{k}{E}\\ \nonumber 
&=&\sum_n\GC{\frac{i}{2k_{ln}^{2l+1}}
\GP{\frac{E^{l+1/2}}{E-k_{ln}^2}}-\frac{i}{2(k_{ln}^*)^{2l+1}}
\GP{\frac{E^{l+1/2}}{E-{k^*_{ln}}^2}}}\\ \nonumber
&=&\pIm{\sum_n\frac{1}{k_{ln}^{2l+1}}\frac{E^{l+1/2}}{k_{ln}^2-E}}, \quad l=0,1,2,\dots
\end{eqnarray}

\section{Survival probability of an unstable state}
For a resonance formed as an intermediate
unstable state in a scattering process, the survival amplitude
(using the Fock-Krylov method) can be expressed as seen before, 
as a Laplace transform of the energy derivative of the scattering phase shift.
Using the Mittag-Leffler expansions derived for the phase shift derivatives 
we shall now try to derive analytical expressions for the survival probabilities and 
discuss their long time behaviour. 

\subsection{General expression for the survival amplitude} 
Replacing Eq. \eqref{E2.86} in \eqref{E2.3}, we get, 
\begin{equation}\label{E2.97}
A_l(t)=
\sum_n{\frac{i}{2k_{ln}}\intseminf{\frac{E^{1/2}}{E-k_{ln}^2}e^{-sE}}{E}}-\sum_n{\frac{i}{2{k_{ln}^*}}\intseminf{\frac{E^{1/2}}{E-{k^*_{ln}}^2}e^{-sE}}{E}}
\end{equation}
If we define $J(\al,\sig,s)$ as
\begin{equation}\label{E2.98}
J(\al,\sig,s)\equiv\intseminf{\frac{E^{\al}}{E+\sig}e^{-sE}}{E},\quad\pRe{s}>0,
\end{equation}
then the survival amplitude can be written as, 
\begin{equation}\label{E2.99}
{
A_l(t)=\sum_n{\frac{i}{2k_{ln}}J\NP{\tfrac{1}{2},-k_{ln}^2,s}}-\sum_n{\frac{i}{2{k_{ln}^*}}J\NP{\tfrac{1}{2},-{k^*_{ln}}^2,s}}}\, .
\end{equation}
In order to ensure that the survival probability, $|A_l(t)|^2$ should be unity at 
$t=0$, we evaluate the normalization factor $A=A_l(t=0)$ and write, 
 \begin{equation}\label{E2.99a}
\boxed{
A_l(t)=\frac{1}{A}\sum_n{\frac{i}{2k_{ln}}J\NP{\tfrac{1}{2},-k_{ln}^2,s}}-\frac{1}{A}\sum_n{\frac{i}{2{k_{ln}^*}}J\NP{\tfrac{1}{2},-{k^*_{ln}}^2,s}}}\, ,
\end{equation}
where $A$ is equal to
\begin{equation}\label{E2.99b}
A=\sum_n{\frac{i}{2k_{ln}}J\NP{\tfrac{1}{2},-k_{ln}^2,c}}-\sum_n{\frac{i}{2{k_{ln}^*}}J\NP{\tfrac{1}{2},-{k^*_{ln}}^2,c}}\, .
\end{equation}
The calculation of the survival amplitude reduces simply to 
the evaluation of the integral 
\footnote{The integral can be found in \cite{Bateman1}, Chapter 4, page 137.}
\eqref{E2.98}. We start with 
\begin{equation}\label{E2.100}
\frac{1}{E+\sig}=\intseminf{e^{-(E+\sig)F}}{F},\quad E>-\pRe{\sig} \, , 
\end{equation}
and replacing in $J$, we get
\begin{align}
J(\al,\sig,s)
&=\intseminf{E^{\al}e^{-sE}\GC{\intseminf{e^{-(E+\sig)F}}{F}}}{E}=
\intseminf{e^{-\sig F}\GC{\intseminf{E^{\al}e^{-(s+F)E}}{E}}}{F}\notag\\
&=\Ga(\al+1)\intseminf{\frac{e^{-\sig F}}{(s+F)^{\al+1}}}{F}=
\Ga(\al+1)e^{\sig s}\int_{s}^{\infty}{\frac{e^{-\sig F}}{F^{\al+1}}}{\,dF}\notag\\
&=\Ga(\al+1)e^{\sig s}\sig^{\al}\int_{\sig s}^{\infty}{\frac{e^{-F}}{F^{\al+1}}}{\,dF}.\label{E2.101}
\end{align}
As a result of the uniform convergence of the 
integral \eqref{E2.100} over the interval of $E$ integration, the order of integration 
can be changed whenever $\pRe{\sig}>0$. 
The last integral can be expressed in terms of the incomplete Gamma functions 
$\Ga(\beta,z)$ \cite{AS,Lebedev}:
\begin{equation}\label{E2.102}
\Ga(\beta,z)=\int_{z}^{\infty}{e^{-t}t^{\beta-1}\,dt},\quad|\pArg{z}|<\pi.
\end{equation}
Therefore, 
\begin{equation}\label{E2.103}
J(\al,\sig,s)=\Ga(\al+1)e^{\sig s}\sig^{\al}\Ga(-\al,\sig s).
\end{equation}
The integral exists if, 
$\pRe{s}>0$, $\pRe{\al}>-1$, $\pRe{\sig}>0$ and $|\pArg{(\sig s)}|<\pi$. 
The last two conditions can be changed to $|\pArg{\sig}|<\pi$. 
Since $\al=l+\frac{1}{2}$, with $l$ being a positive integer, 
it is possible to express $J$ in terms of error functions. 
The latter is very useful in the evaluation of the survival amplitudes numerically. 
In order to do this, we begin with 
\begin{equation}\label{E2.104}
\frac{E^{l}}{E+\sig}=\sum_{p=0}^{l-1}{(-1)^p\sig^{p}E^{l-p-1}}+(-1)^l\frac{\sig^l}{E+\sig}.
\end{equation}
Replacing in the definition of $J$, we get
\begin{align}
J\NP{l+\tfrac{1}{2},\sig,s}
&=\sum_{p=0}^{l-1}{(-1)^p\sig^{p}\intseminf{E^{l-p-1/2}e^{-sE}}{E}}+
(-1)^l\sig^l\intseminf{\frac{E^{1/2}}{E+\sig}e^{-sE}}{E}\notag\\
&=(-1)^l\sig^{l+1/2}\EC{\,\sum_{p=0}^{l}{(-1)^p\frac{\Ga(p+\frac{1}{2})}{\sig^{l+1/2}s^{l+1/2}}}-\pi e^{\sig s}\erfc{\NP{\sig^{1/2}s^{1/2}}}},\quad\label{E2.105}
\end{align}
where $|\pArg{\sigma}|<\pi$ and $\pRe{s}>0$. The above was done by using the following integral \cite{Bateman1}:
{\[
\intseminf{\frac{E^{1/2}}{E+\sig}e^{-sE}}{E}=\sqrt{\frac{\pi}{s}}-\pi\sig^{1/2}e^{\sig s}\erfc{\NP{\sig^{1/2}s^{1/2}}},\quad|\pArg{\sigma}|<\pi,\pRe{s}>0.
\]}
Eq. \eqref{E2.105} can also be deduced from \eqref{E2.103} using recurrence relations of 
the incomplete gamma functions. This allows us to write the results as follows\footnote{An expression for the survival amplitude of a Breit-Wigner resonance, 
in terms of hypergeometric functions is derived in \cite{Brzeski}.}: 
\begin{equation}\label{E2.106}
\boxed{
A_l(t)=-\frac{1}{2A}\Ga\NP{\tfrac{3}{2}}\sum_n{\EC{e^{-k_{ln}^2s}\Ga\NP{-\tfrac{1}{2},-k_{ln}^2s}
-e^{-{k_{ln}^*}^2s}\Ga\NP{-\tfrac{1}{2},-{k_{ln}^*}^2s}}}}\,.
\end{equation}
Also, the factor $A$ can be calculated in a similar way:
\begin{equation}\label{E2.106b}
{
A=-\frac{1}{2}\Ga\NP{\tfrac{3}{2}}\sum_n{\EC{e^{-k_{ln}^2c}\,\Ga\NP{-\tfrac{1}{2},-k_{ln}^2c}
-e^{-{k_{ln}^*}^2c}\,\Ga\NP{-\tfrac{1}{2},-{k_{ln}^*}^2c}}}}\,.
\end{equation}
The above $A_l(t)$ gives us a general, model independent expression for the 
survival amplitude of resonances occuring in the $l^{th}$ partial wave.

\subsection{Large time behaviour of the survival probability} 
It is well known by now that the exponential decay is only an approximation
and quantum mechanics predicts a non-exponential decay (actually power laws) 
at very short and
large times \cite{urbanowski} in the time evolution of an unstable state.
Hence, using the results obtained so far, we shall now examine the behaviour
of the survival amplitude at large times. In order to examine the survival 
probability at large times, we begin with \cite{Copson2} 
\[
e^{z}\Ga\NP{-\al,z}\sim\sum_{r=0}^{\infty}{(-1)^r\frac{\Ga(\al+1+r)}{\Ga(\al+1)}\,z^{-(\al+1+r)}},\quad \al>0.
\]
Replacing the above expression in Eq. \eqref{E2.106}, one obtains 
\begin{align}
A_l(t)
&\sim-\frac{1}{2A}\sum_n{\sum_{r=0}^{\infty}{(-1)^r\Ga\NP{r+\tfrac{3}{2}}\EC{\frac{1}{\NP{-k_{ln}^2s}^{r+3/2}}-\frac{1}{\NP{-{k_{ln}^*}^2s}^{r+3/2}}}}}\notag\\
&\sim\frac{1}{A}\sum_{r=0}^{\infty}{\frac{\Ga\NP{r+\tfrac{3}{2}}}{s^{r+3/2}}\pIm{\GP{\sum_n{\frac{1}{k_{ln}^{2r+3}}}}}}.\label{E2.107}
\end{align}
Using the condition \eqref{E2.88}, and noting that $s\sim it$, we have 
\begin{equation}\label{E2.108}
A_l(t)\sim\frac{1}{A}\sum_{r=l}^{\infty}{\frac{\Ga\NP{r+\tfrac{3}{2}}}{i^{r+3/2}}\pIm{\GP{\sum_n{\frac{1}{k_{ln}^{2r+3}}}}}\,t^{-r-3/2}}.
\end{equation}
The dominant contribution is given by the term for which $r=l$. Therefore,
\begin{equation}\label{E2.109}
A_l(t)\sim\frac{1}{A}{\frac{\Ga\NP{l+\tfrac{3}{2}}}{i^{l+3/2}}\pIm{\GP{\sum_n{\frac{1}{k_{ln}^{2l+3}}}}}\,t^{-l-3/2}},
\end{equation}
and the survival probability, $P_l(t) =|A_l(t)|^2$ at large times is proportional 
to $t^{-2l-3}$. This is indeed in accordance with literature \cite{ournonexpo1,fonda}. 

\subsection{Survival amplitude for a Breit-Wigner resonance plus corrections}
If we begin with a Breit-Wigner distribution for the density of states, the 
survival amplitude would be proportional to 
\begin{equation}\label{E2.140}
A_l^{BW}(t)\propto\intseminf{\frac{E^{1/2}e^{-cE}e^{-iEt}}{(E-\ep_{lr})^2+(\Ga_{lr}/2)^2}}{E}.
\end{equation}
Since $k_{lr}^2=\ep_{lr}-i\Ga_{lr}/2$, the above integral can be written as 
\begin{equation}\label{E2.141}
\intseminf{\frac{E^{1/2}e^{-sE}}{(E-k_{lr}^2)(E-{k_{lr}^*}^2)}}{E}=
\frac{i}{\Ga_{lr}}\intseminf{\GP{\frac{E^{1/2}e^{-sE}}{E-k_{lr}^2}-\frac{E^{1/2}e^{-sE}}{E-{k_{lr}^*}^2}}}{E}.
\end{equation}
Expressing it further in terms of $J$, one obtains 
\begin{equation}\label{E2.142}
A_l^{BW}(t)\propto\MC{J\NP{\tfrac{1}{2},-k_{lr}^2,s}-J\NP{\tfrac{1}{2},
-{k_{lr}^*}^2,s}}.
\end{equation}
Thus the survival amplitude for a Breit-Wigner resonance would be proportional to the 
difference of the functions 
$J\NP{\tfrac{1}{2},-k_{lr}^2,s}$ and $J\NP{\tfrac{1}{2},-{k_{lr}^*}^2,s}$. 
The latter suggests that the above result can also be obtained if we start with 
the general expression obtained earlier if we expand $k_{ln}^{-1}$ and 
${k_{ln}^*}^{-1}$ in powers of $x_{ln}=\Ga_{ln}/{2\ep_{ln}}$. 
From the expansion 
\begin{equation}\label{E2.143}
(1+z)^{\nu}=\Ga\NP{\nu+1}\sum_{q=0}^{\infty}{\frac{z^q}{q!\Ga\NP{\nu+1-q}}},
\quad \nu\neq-1,-2,\dotsc
\end{equation}
with $\nu=-1/2$ and $z=x_{ln}$, we can write $k_{ln}^{-1}$ and ${k_{ln}^*}^{-1}$ as
\begin{align}
\frac{1}{k_{ln}}&=\frac{1}{\sqrt{\ep_{ln}}}\NP{1-ix_{ln}}^{-1/2}=\frac{1}{\sqrt{\ep_{ln}}}\,\Ga\NP{\tfrac{1}{2}}\sum_{q=0}^{\infty}{\frac{e^{-iq\pi/2}}{q!\Ga\NP{\tfrac{1}{2}-q}}\,x_{ln}^q},\label{E2.144}\\
\frac{1}{k_{ln}^*}&=\frac{1}{\sqrt{\ep_{ln}}}\NP{1+ix_{ln}}^{-1/2}=\frac{1}{\sqrt{\ep_{ln}}}\,\Ga\NP{\tfrac{1}{2}}\sum_{q=0}^{\infty}{\frac{e^{\,iq\pi/2}}{q!\Ga\NP{\tfrac{1}{2}-q}}\,x_{ln}^q},\label{E2.145}
\end{align}
Replacing the above in Eq. \eqref{E2.99a}, we obtain 
\begin{equation}\label{E2.146}
A_l(t)=\frac{i}{2A}\sum_{q=0}^{\infty}{\frac{\Ga\NP{\tfrac{1}{2}}}{q!\Ga\NP{\tfrac{1}{2}-q}}\sum_{n}{\frac{1}{\sqrt{\ep_{ln}}}\,\MC{e^{-iq\pi/2}J\NP{\tfrac{1}{2},-k_{ln}^2,s}-e^{\,iq\pi/2}J\NP{\tfrac{1}{2},-{k_{ln}^*}^2,s}}\,x_{ln}^q}}.
\end{equation}
Writing the first few terms, the survival amplitude is given as, 
\begin{eqnarray}\label{E2.147}
A_l(t)&=&\frac{i}{2A}\sum_{n}{\frac{1}{\sqrt{\ep_{ln}}}\,
\MC{J\NP{\tfrac{1}{2},-k_{ln}^2,s}-J\NP{\tfrac{1}{2},-{k_{ln}^*}^2,s}}}\\ \nonumber
&+&\frac{i}{2A}\sum_{n}{\frac{i}{2\sqrt{\ep_{ln}}}\,\MC{J\NP{\tfrac{1}{2},-k_{ln}^2,s}+J\NP{\tfrac{1}{2},-{k_{ln}^*}^2,s}}\GP{\frac{\Ga_{ln}}{2\ep_{ln}}}}\\ \nonumber
&-&\frac{i}{2A}\sum_{n}{\frac{3}{8\sqrt{\ep_{ln}}}\,\MC{J\NP{\tfrac{1}{2},-k_{ln}^2,s}-
J\NP{\tfrac{1}{2},-{k_{ln}^*}^2,s}}\GP{\frac{\Ga_{ln}}{2\ep_{ln}}}^2}\\ \nonumber
&+&\frac{i}{2A}\sum_{n}{\frac{5i}{16\sqrt{\ep_{ln}}}\,\MC{J\NP{\tfrac{1}{2},-k_{ln}^2,s}+J\NP{\tfrac{1}{2},-{k_{ln}^*}^2,s}}\GP{\frac{\Ga_{ln}}{2\ep_{ln}}}^3}+\cdots
\end{eqnarray}
Note that the first term here is similar to \eqref{E2.142} and the next terms 
are corrections suppressed by powers of $\Gamma_{ln}/2\epsilon_{ln}$. 

\section{Applications and limitations}
In the previous sections, we derived analytical, model independent expressions for 
phase shifts, energy derivatives of phase shift (which can be related to the 
density of states in a resonance) and cross sections in single channel elastic 
scattering solely by using the analyticity properties of the $S$-matrix and 
a theorem of Mittag-Leffler. The latter also allowed us to derive the survival 
amplitudes with the Fock-Krylov method and confirm their 
quantum mechanical behaviour which is non-exponential at large times.
We shall now apply all the above ``exact" expressions to realistic examples. 
However, it is important to note that within the formalism developed here, we 
cannot be too ambitious in reproducing the experimental results due to the 
following reasons:\\
(i) the present formalism restricts to elastic scattering and most 
(hadronic) resonances 
can decay to multiple channels. The $S$-matrix considered here does not take 
into account the effect of inelasticities.\\
(ii) Non-resonant background can distort the shape of the 
purely resonant cross sections and the extracted phase shifts 
may not necessarily be described simply by the pole structure (see for example
\cite{arndt3} where the $S$-matrix is parametrized with a resonant and 
non-resonant part).\\
Nevertheless, in what follows, we shall try to test how far the present approach 
succeeds within the limitations mentioned above.

\subsection{Low energy neutron-proton scattering}
We start with a pedagogical example of the total cross section for 
low energy neutron proton (np) scattering where one is aware of the existence 
of the bound state, namely, the deuteron with a binding energy of 2.2245 MeV. 
It is also well known that the total np cross section cannot be reproduced 
unless one considers the existence of a virtual (singlet) state \cite{schwinger} 
around 0.1 MeV below the np threshold.
Beginning with the cross section expression derived in this work and including 
one bound state and one virtual state pole, 
we evaluate the cross section  for np scattering 
with $l =0$.
The deuteron in principle is a spin 1 nucleus and can be described as 
an admixture of the $^3$S$_1$-$^3$D$_1$ states. The virtual state is $^1$S$_0$. 
Without entering into the details of the amount of the above admixture 
(which can range between 3 - 7 \% \cite{mePLB}) and nuclear physics aspects, 
we attempt to calculate the np total cross section using the above formula for 
$s$-wave scattering. In Fig. 2, we see the results of the calculations as 
compared to data \cite{nndcdata}. 
\begin{figure}[h]
\includegraphics[width=12cm,height=8cm]{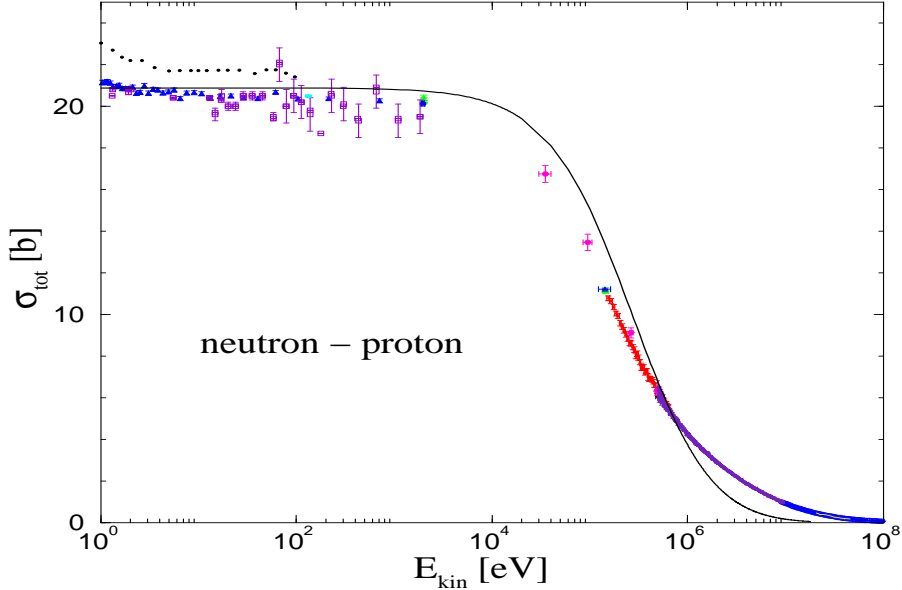}
\caption{Comparison of the theoretical cross section expression with the 
neutron-proton low energy total cross section data \cite{nndcdata}.}
\end{figure}
The agreement with data is reasonable and indeed the magnitude of the 
cross section is sensitive to the existence of the virtual state as expected. 
The theoretical curve shown in Fig. 2 is evaluated using (\ref{E2.126}) with 
a bound state at -2.2245 MeV and a virtual state at -0.14 MeV. 

\subsection{Phase shifts in pion nucleon scattering}
Our next example is that of scattering phase shifts extracted from the data 
on pion nucleon ($\pi$N)scattering. In Fig. 3, we present the single energy values 
\cite{wetimedelay} extracted from a partial wave analysis of the $\pi$N 
data corresponding to the resonance regions of the $\Delta$(1232) and 
N(1520) resonances in the P$_{33}$ and D$_{13}$ (notation: $l_{2I,2J}$) 
waves respectively. The agreement of the calculated phase shifts with the 
single energy values is not very good. However, the energy derivative of the 
phase shift does reproduce the peak structure at the pole position. 
The agreement was not expected to be very good due to the reasons 
(such as inelasticities and non-resonant processes) already mentioned before. 
\begin{figure}[h]
\includegraphics[width=12cm,height=8cm]{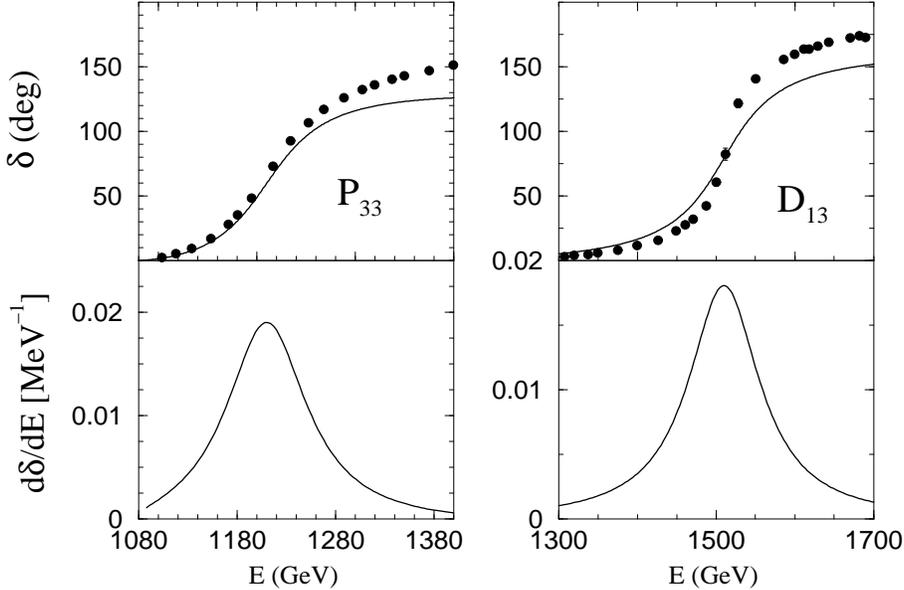}
\caption{Comparison of the calculated 
phase shifts with the single energy values given in \cite{wetimedelay}.}
\end{figure}

\subsection{Survival probabilities and critical times}
The decay law of an unstable particle (or nucleus) can be shown classically 
to be of an exponential nature but according to quantum mechanics, 
this law is an approximation which fails for short and large times
\cite{ournonexpo1,ournonexpo2}.
The non-exponential behaviour predicted by quantum mechanics has intrigued
experimental nuclear and particle physicists who performed experiments
(see \cite{norman} and references therein) with
nuclei such as $^{222}$Rn, $^{60}$Co, $^{56}$Mn and measured the decay law
for several half-lives to disappointingly find only an exponential decay law.

Using the expressions derived in the present work, we shall now evaluate the 
decay law (the survival probability) for a broad and a narrow resonance and 
try to understand the possible reasons behind not observing the non-exponential 
decay law at large times. We consider the case of the $\sigma$ meson as a 
typical broad resonance and that of the lowest excited state of the 
$^8$Be nucleus as the narrow resonance. 
In Fig. 4 we notice the typical exponential decay law
for $^8$Be (which has 100 \% decay to two $\alpha$'s), followed by an 
oscillatory transition 
region (see inset) and a power law at large times. The broad resonance does not show the 
same characteristics. Indeed, it does not show an exponential behaviour at all 
(consistent with previous theoretical calculations \cite{ournonexpo2}) at any time. 
We show the behaviour for 4 different pole values taken from literature \cite{pdg}. 
Though the non-exponential behaviour in the beginning is different for the different 
pole values, the power law sets in at roughly the same time and all curves exhibit 
the same power law behaviour of $t^{-3}$ (as expected for $l =0$). 
\begin{figure}[h]
\includegraphics[width=14cm,height=8cm]{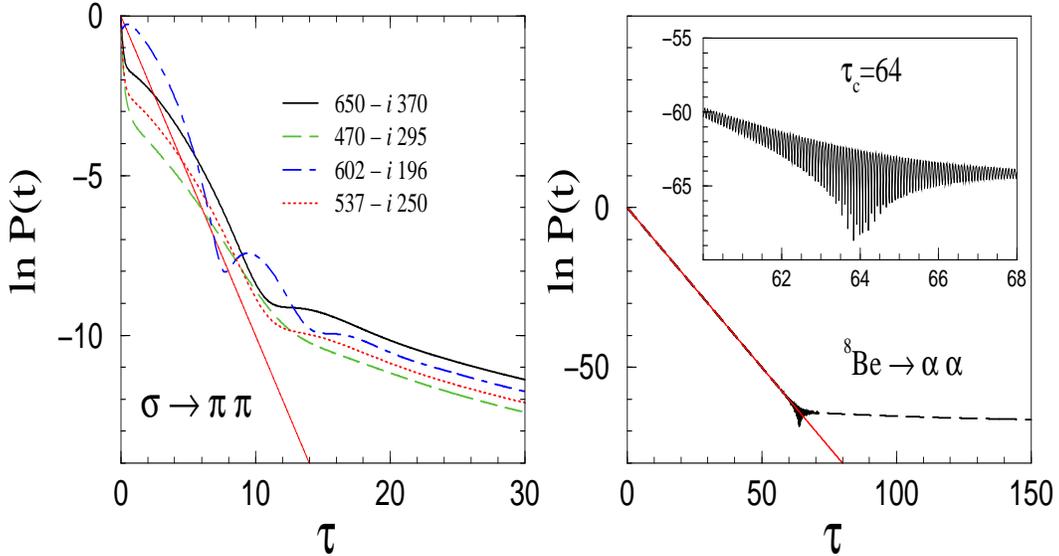}
\caption{Survival probability of the $\sigma$ meson (left) and the $^8$Be nucleus
(right).}
\end{figure}

The number of half-lives after which the power law starts for the $^8$Be(0$^+$) 
resonance is about 60. A similar calculation for a narrow nuclear resonance such as 
$^{56}$Mn leads to a critical time of 300 half-lives for the 
power law to set in. 
The experiment in case of $^{56}$Mn \cite{norman} 
was carried out only up to 45 half-lives and the non-exponential was not observed. 
However, performing measurements up to 300 half-lives would be practically
impossible since the exponential decay law would destroy almost all
the sample by the time the narrow resonance reaches the power law.
Broad resonances such as the $\sigma$ meson reach the power law much
earlier, however, the lifetime is too short making the experimental
observation once again difficult.

\section{Summary and Conclusions}
Performing a Mittag-Leffler (ML) based analysis of the $S$-matrix and 
related quantities such as the phase shifts and cross sections, it is shown that 
a knowledge of the poles is sufficient to determine these quantities. 
We summarize the main findings of the present work below:
\begin{itemize}
\item[1.] The ML expansion of the $S$-matrix in terms of its poles 
and the corresponding residues, along with the conditions imposed 
due to the analytic properties of the $S$-matrix is presented. Analytic 
expressions for the residues are derived.
\item[2.] Making use of the optical theorem
and the ML expansion of the $S$-matrix, expressions for the cross sections with the  
corresponding poles are derived. Restricting these expressions to 
the case of resonant poles, generalized formulae for resonances in single channel 
elastic scattering are obtained.
\item[3.] The generalized cross sections are shown to reduce to the standard 
Breit-Wigner (BW) cross sections plus corrections. The BW formula 
overestimates the cross sections. However, with the examples of a 
broad meson and a baryon resonance, we show that adding the correction terms, the cross 
sections decrease and increase alternately, getting closer to the exact result.  
The number of correction terms required depends on the ratio, $\Gamma_r/(2\epsilon_r)$, 
of the half width and the energy ($\epsilon_r = E_r - E_{th}$) above threshold. 
\item[4.] Performing the 
Mittag-Leffler based analysis with a focus on the correct threshold behaviour, 
we find that additional conditions must be imposed on the residues. 
It is gratifying to find that the generalized formulae near threshold also reduce to 
the Breit-Wigner forms plus corrections.
\item[5.] A Mittag-Leffler based analysis of the energy derivative of the 
phase shift is also presented and further used in order to derive analytical 
expressions 
for the survival amplitudes of resonances. Thus, it is shown that a knowledge 
of the pole value is sufficient to calculate the survival probability of an 
unstable state analytically within the Fock-Krylov method.
The large time (power law) behaviour of these survival amplitudes is consistent 
with expectations from literature.
\item[6.] Applications of the analytical expressions derived are presented in the 
last section. Calculations of the survival probabilities of broad and narrow resonances 
shed some light on the reason for the non-observability of the non-exponential decay
law predicted by quantum mechanics at large times.
\end{itemize}

Having summarized the main findings, we must add a few words of caution. 
The present work is a first step in the derivation of 
non-relativistic cross sections and survival 
probabilities in a model independent way and hence deals only with the case of 
single channel elastic scattering. A realistic scattering amplitude 
must of course take into 
account the effects due to multiple channels and background processes. 
Apart from this, the non-relativistic framework may not be the best as far 
as hadronic resonances are concerned.
However, a relativistic extension to obtain analytical formulae as in the present work 
can be quite challenging \cite{abohmsato,abohmmithai}. 
For a more realistic approach, we refer the reader to 
\cite{workmanplb,workmanprc}, where, using the Laurent (Mittag-Leffler) expansion, 
the authors fitted data to extract the pole positions and residues. 
We hope to perform an extension to consider the above points in future.

\begin{acknowledgments}
The authors thank M. Nowakowski for a careful reading of the manuscript and 
his critical remarks and suggestions. One of the authors (N. G. K.) acknowledges the 
support from the Faculty of Science, Universidad de los Andes, Colombia, 
through grant no. P18.160322.001-17.
\end{acknowledgments}
\appendix

\section{Unimodularity of the Mittag-Leffler expansion of the $S$-matrix}
The $S$-matrix, when $k$ is real, is unimodular \cite{Baz}:
\begin{equation}\label{E2.24}
S_l(k)S_l^*(k)=1.
\end{equation}
From \eqref{E2.18}, we can see that the poles and zeros of the $S$-matrix 
are all poles of $S_l(k)S_l^*(k)$. Thus, it is possible to apply the ML 
theorem to this function. 
The value of the latter at $k=0$ is 1, and its residues are computed with the help 
of \eqref{E2.22} and \eqref{E2.23}:
\begin{enumerate}[a)]
\item Residue at $k=k_{ln}$:
\begin{align*}
&\Res\NC{{S_l(k)S_l^*(k),k=k_{ln}}}=\\&b_{ln}\EC{1+k_{ln}\sum_p{\GC{\frac{b_{lp}^*}{k_{lp}^*(k_{ln}-k_{lp}^*)}+\frac{b_{lp}}{k_{lp}(k_{ln}+k_{lp})}}}+k_{ln}\sum_q{\frac{D_{lq}}{\zeta_{lq}(k_{ln}+i\zeta_{lq})}}}=0.
\end{align*}
\item Residue at $k=-k_{ln}^*$:
\begin{align*}
&\Res\NC{{S_l(k)S_l^*(k),k=-k_{ln}^*}}=\\&b_{ln}^*\EC{1+k_{ln}^*\sum_p{\GC{\frac{b_{lp}}{k_{lp}(k_{ln}+k_{lp})}+\frac{b_{lp}^*}{k_{lp}^*(k_{ln}-k_{lp}^*)}}}+k_{ln}\sum_q{\frac{D_{lq}}{\zeta_{lq}(k_{ln}+i\zeta_{lq})}}}^*=0.
\end{align*}
\item Residue at $k=-k_{ln}$:
\begin{align*}
&\Res\NC{{S_l(k)S_l^*(k),k=-k_{ln}}}=\\&-b_{ln}\EC{1+k_{ln}\sum_p{\GC{\frac{b_{lp}}{k_{lp}(k_{ln}+k_{lp})}+\frac{b_{lp}^*}{k_{lp}^*(k_{ln}-k_{lp}^*)}}}+k_{ln}\sum_q{\frac{D_{lq}}{\zeta_{lq}(k_{ln}+i\zeta_{lq})}}}=0.
\end{align*}
\item Residue at $k=k_{ln}^*$:
\begin{align*}
&\Res\NC{{S_l(k)S_l^*(k),k=k_{ln}^*}}=\\&b_{ln}^*\EC{1+k_{ln}\sum_p{\GC{\frac{b_{lp}^*}{k_{lp}^*(k_{ln}-k_{lp}^*)}+\frac{b_{lp}}{k_{lp}(k_{ln}+k_{lp})}}}+k_{ln}\sum_q{\frac{D_{lq}}{\zeta_{lq}(k_{ln}+i\zeta_{lq})}}}^*=0.
\end{align*}
\item Residue at $k=i\zeta_{lm}$:
\begin{align*}
&\Res\NC{{S_l(k)S_l^*(k),k=i\zeta_{lm}}}=\\&iD_{lm}\EC{1+i\zeta_{lm}\sum_p{\GC{\frac{b_{lp}}{k_{lp}(i\zeta_{lm}-k_{lp})}+\frac{b_{lp}^*}{k_{lp}^*(i\zeta_{lm}+k_{lp}^*)}}}+\zeta_{lm}\sum_q{\frac{D_{lq}}{\zeta_{lq}(\zeta_{lm}-\zeta_{lq})}}}=0.
\end{align*}
\item Residue at $k=-i\zeta_{lm}$:
\begin{align*}
&\Res\NC{{S_l(k)S_l^*(k),k=-i\zeta_{lm}}}=\\&-iD_{lm}\EC{1+i\zeta_{lm}\sum_{p}{\GC{\frac{b_{lp}^*}{k_{lp}^*(i\zeta_{lm}+k_{lp}^*)}+\frac{b_{lp}}{k_{lp}(i\zeta_{lm}-k_{lp})}}}+\zeta_{lm}\sum_q{\frac{D_{lq}}{\zeta_{lq}(\zeta_{lm}-\zeta_{lq})}}}=0.
\end{align*}
\end{enumerate}
Since all residues of $S_l(k)S_l^*(k)$ are zero, from the ML theorem, 
$S_l(k)S_l^*(k)=1$ and hence we can say that Eq. \eqref{E2.18} satisfies the 
unimodularity condition.

\section{Threshold conditions for the Mittag-Leffler expansions}
We want to study the behavior of a function with a zero (not necessarily simple) 
at the origin of the complex momentum plane through its ML expansion. 
The $S$-matrix, total cross section and density of states are examples of such 
functions. Let $g(z)$ be the function which satisfies the ML theorem. 
Also, $g(z)$ has a zero of multiplicity $m$ at $z=0$. Consider the integral
\begin{equation}\label{E2.111}
\frac{1}{2\pi i}\oint_{C_N}{\frac{g(\zeta)}{\zeta^{m}(\zeta-z)}\,d\zeta},
\end{equation}
where $z\neq a_n$ is different from zero and is inside the contour of integration $C_N:|\zeta|=R_N$, which contains $N$ poles of $g(\zeta)$. From its definition, 
this radius is such that
\begin{equation}\label{E2.112}
|a_N|<R_N<|a_{N+1}|.
\end{equation} 
The integrand has simple poles at $\zeta=a_n$ and $\zeta=z$. 
Using the residue theorem, we get:
\begin{equation}\label{E2.113}
\frac{1}{2\pi i}\oint_{C_N}{\frac{g(\zeta)}{\zeta^{m}(\zeta-z)}\,d\zeta}=\frac{g(z)}{z^m}+\sum_{n=1}^{N}{\frac{b_n}{a_n^m(a_n-z)}},
\end{equation} 
where the sum is over all poles closed by $C_N$. 
Since $|g(z)|<M$ on $C_N$, with $M$ constant,
\begin{equation}\label{E2.114}
\EB{\frac{1}{2\pi i}\oint_{C_N}{\frac{g(\zeta)}{\zeta^{m}(\zeta-z)}\,d\zeta}}\leq\frac{1}{2\pi}\frac{M\cdot 2\pi R_N}{R_N^m\NP{R_N-|z|}}=\cfrac{M}{R_N^m\GP{1-\cfrac{|z|}{R_N}}}.
\end{equation}  
If $N\to\infty$, $R_N\to\infty$ and the integral tends to zero if $m>0$. Thus,
\begin{equation}\label{E2.115}
\frac{g(z)}{z^m}=\sum_{n=1}^{\infty}{\frac{b_n}{a_n^m(z-a_n)}}.
\end{equation}
If $z\to0$,
\begin{equation}\label{E2.116}
\lim_{z\to0}{\GB{\frac{g(z)}{z^m}}}=\GB{\sum_{n=1}^{\infty}{\frac{b_n}{a_n^{m+1}}}}\leq\sum_{n=1}^{\infty}{\GB{\frac{b_n}{a_n^{m+1}}}}.
\end{equation}
If the last series converges absolutely, $g(z)=O(z^m)$ for $z\to 0$. 
On the other hand, if $z\to\infty$,
\begin{equation}\label{E2.117}
\lim_{z\to\infty}{\GB{\frac{g(z)}{z^m}}}=0.
\end{equation}
We have, for $z\to\infty$, $g(z)=o(z^m)$. 

If we apply the ML theorem to $g(z)$:
\begin{equation}\label{E2.7a}
g(z)=\sum_{n=1}^{\infty}{{\frac{b_nz}{a_n(z-a_n)}}},
\end{equation}
and we compare this expansion with \eqref{E2.115}, it is natural to ask ourselves how 
the two are connected; what we mean is, 
how do we deduce \eqref{E2.115} from \eqref{E2.7a}? 
The key is to expand in a Taylor series \eqref{E2.7a} and see what conditions 
should we impose on the residues and poles of $g(z) $ for obtaining the zero at 
$z=0$ with the correct multiplicity. 

Since
\begin{equation}\label{E2.119}
\frac{1}{z-a_n}=-\frac{1}{a_n}\sum_{s=0}^{\infty}{\GP{\frac{z}{a_n}}^s},\quad|z|<|a_n|,
\end{equation}
thus
\begin{equation}\label{E2.120}
g(z)=-\sum_{n=1}^{\infty}{\GP{\frac{b_n}{a_n}}\sum_{s=0}^{\infty}{\GP{\frac{z}{a_n}}^{s+1}}}
=-\sum_{s=1}^{\infty}{z^{s}\GP{\sum_{n}{\frac{b_n}{a_n^{s+1}}}}},\quad|z|<\min_{n}{|a_n|}=|a_1|.
\end{equation}
The first $m-1$ terms of the last expansion must be zero, 
since $g(z)$ has a zero at $z=0$ of order $m$. This means that
\begin{equation}\label{E2.121}
\sum_{n}{\frac{b_n}{a_n^{s+1}}}=0,\quad s=1,2,\dotsc,m-1.
\end{equation}
Conditions \eqref{E2.121} provide a bridge between the expansions \eqref{E2.7a} and 
\eqref{E2.115} in the following way: from identity
\footnote{This can be deduced 
if we write $\dfrac{1}{z-w}$ as:
\[
\frac{1}{z-w}=\frac{w^{n}-z^n+z^n}{w^n(z-w)}=\GP{\frac{z}{w}}^n\frac{1}{z-w}-\GP{\frac{1}{w}}\cfrac{1-\NP{z/w}^n}{1-z/w}=
\GP{\frac{z}{w}}^n\frac{1}{z-w}-\frac{1}{w}\sum_{s=0}^{n-1}{\GP{\frac{z}{w}}^s}.
\]
Here, $n\geq1$.}
\begin{equation}\label{E2.122}
\frac{1}{z-w}=\GP{\frac{z}{w}}^p\frac{1}{z-w}-\frac{1}{w}\sum_{s=0}^{p-1}{\GP{\frac{z}{w}}^s},\quad p=1,2,\dotsc
\end{equation}
with $w=a_n$ and $p=m-1$; we have $f(z)$, given by \eqref{E2.7}, is equal to
\begin{align}
g(z)
&=\sum_{n}{b_n\,\frac{z}{a_n}\GC{\GP{\frac{z}{a_n}}^{m-1}\frac{1}{z-a_n}-\frac{1}{a_n}\sum_{s=0}^{m-2}{\GP{\frac{z}{a_n}}^s}}}\notag\\
&=z^m\sum_{n}{\frac{b_n}{a_n^{m}\NP{z-a_n}}}-\sum_{s=0}^{m-2}{z^{s+1}\GP{\sum_n{\frac{b_n}{a_n^{s+2}}}}}\notag\\
&=z^m\sum_{n}{\frac{b_n}{a_n^{m}\NP{z-a_n}}}-\sum_{s=1}^{m-1}{z^{s}\GP{\sum_n{\frac{b_n}{a_n^{s+1}}}}}.\label{E2.123}
\end{align}
Using \eqref{E2.121}, we obtain the expansion \eqref{E2.115}.

\end{document}